\documentclass[11pt]{amsart}

\usepackage{amsmath}
\usepackage{amsthm}
\usepackage{amssymb}
\usepackage{amsfonts}
\usepackage[round]{natbib}
\usepackage{subfig}
\usepackage[foot]{amsaddr}
\usepackage{booktabs}
\usepackage{multirow}
\usepackage{algorithm2e}
\RestyleAlgo{ruled}

\newcommand{\ring}[1]{\ensuremath{\mathbb{#1}}}

\newcommand\NN{\ring{N}}

\newcommand\RR{\ring{R}}

\newcommand\ZZ{\ring{Z}}

\newcommand\cM{{\mathcal M}}

\newcommand\cP{{\mathcal P}}

\newcommand\cW{{\mathcal W}}
\newcommand{\fiber}[2]{\mathcal{F}_{#1,#2}}
\newcommand{\fibergraph}[3]{\fiber{#1}{#2}\left(#3\right)}
\newcommand{\metropolis}[2]{\mathcal{M}_{#1,#2}}

\newcommand{\reject}[3]{p_{#1}\left(#2, #3\right)}

\usepackage{hyperref}
\usepackage{url}
\usepackage{pgfplots}
\pgfplotsset{compat=1.18}

\usepackage{tkz-berge}

\usetikzlibrary{patterns}
\usetikzlibrary{positioning}
\usetikzlibrary{decorations.pathreplacing} 
\usetikzlibrary{graphs,graphs.standard}
\usetikzlibrary{calc}
\usetikzlibrary{angles}
\usetikzlibrary{matrix}
\usetikzlibrary{fadings}
\usetikzlibrary{backgrounds}

\GraphInit[vstyle=Classic]
\SetGraphUnit{1}
\SetVertexMath
\tikzset{EdgeStyle/.style = {bend right, font=\scriptsize}} 
\tikzset{VertexStyle/.style = {shape = circle,fill = black,minimum size = 1pt, node font=\scriptsize}}

\newtheorem{thm}{Theorem}[section]
\newtheorem{lemma}[thm]{Lemma}

\theoremstyle{definition}

\numberwithin{equation}{section}

\title{SAT-sampling for statistical significance testing in sparse contingency tables}

\author{Patrick Scharpfenecker$^\ast$}
\author{Tobias Windisch}
\thanks{Both authors are with the University of Applied Sciences Kempten, Germany.} 
\email{$\{\texttt{patrick.scharpfenecker}, \texttt{tobias.windisch}\}\texttt{@hs-kempten.de}$}

\newcommand{\hybridAlg}{\mathrm{A}_{n}(\cM)}
\newcommand{\parallelAlg}{\mathrm{P}_{n, k}(\cM)}

\begin{document}

\begin{abstract}
Exact conditional tests for contingency tables require sampling from fibers with fixed margins.
Classical Markov basis MCMC is general but often impractical: computing full Markov bases that
connect all fibers of a given constraint matrix can be infeasible
and the resulting chains may converge slowly, especially in sparse settings or in presence of
structural zeros. We introduce a SAT-based alternative that encodes fibers as Boolean circuits which allows
modern SAT samplers to generate tables randomly. We analyze the
sampling bias that SAT samplers may introduce, provide diagnostics, and propose practical mitigation. 
We propose hybrid MCMC schemes that combine SAT proposals with local moves to ensure correct
stationary distributions which do not necessarily require connectivity via local moves which is
particularly beneficial in presence of structural zeros.
Across benchmarks, including small and involved tables with many structural zeros where pure
Markov-basis methods underperform, our methods deliver reliable conditional p-values and often
outperform samplers that rely on precomputed Markov bases.
\end{abstract}

\maketitle

\begingroup
\renewcommand\thefootnote{}%
\footnotetext{$^\ast$ Corresponding author.}%
\endgroup

\keywords{Metropolis Hastings; Markov Chain Monte Carlo; Independence Testing; Satisfiability}

\maketitle

\section{Introduction}

Exact conditional inference for discrete exponential families, such as independence testing in
contingency tables, hinges on computations over \emph{fibers}—the sets of nonnegative integer tables
with fixed sufficient statistics. A standard approach is to construct an irreducible Markov chain
over a fiber using a Markov basis and to approximate conditional $p$-values by Markov chain Monte
Carlo (MCMC). The use of Markov-basis–style moves and exact conditional logic has recently extended
to domains such as neural spatial interaction models~\cite{NEURIPS2024_c7313c42}, underscoring 
broader relevance of scalable sampling over constrained integer lattices. 

While widely applicable in theory, this strategy faces two well-known bottlenecks in
practice. First, computing a full Markov basis \emph{a priori} is often computationally prohibitive
or outright infeasible for realistic models and table sizes~\cite{no3f_3kk}. Second, even when a
basis is available, the induced moves can mix slowly and require substantial tuning
effort~\cite{RapidMixingMarkovBases}. These issues are exacerbated by structural zeros and other
constraints, which can inflate the size of Markov bases and complicate
connectivity~\cite{rapallo2006markov}.
To mitigate these challenges, several lines of work have aimed to reduce—and sometimes eliminate—the
dependence on precomputed Markov bases. 
Some works leverage
lattice-basis moves and study when such bases suffice for connectivity~\cite{lattice_basis_moves},
or develop samplers that forgo Markov bases entirely in favor of more direct lattice-based
proposals~\cite{sampling_without_markov_basis}. More recently, RUMBA introduced randomized updates
that sample lattice points efficiently in high dimensions~\cite{bakenhusSamplingLatticePoints2023}.
In parallel, problem-specific strategies have been proposed for independence testing in large and
sparse tables~\cite{indep_large_sparse_cts}. Collectively, these methods trade off generality, ease
of implementation, and statistical efficiency, and they highlight a core need: flexible, scalable
samplers that deliver reliable conditional $p$-values without incurring the full cost of
Markov-basis computation.

\begin{figure}
\begin{center}
    
\subfloat[]{
\begin{tikzpicture}[
    scale=0.5,
    polynode/.style={draw=black, circle, fill=black, inner sep=2pt}
    ]

\begin{axis}[
  width=8cm, height=6cm,
  axis equal image,        
  xmin=0, xmax=8,
  ymin=0, ymax=6,
  grid=both,
  minor tick num=1,
  xlabel={}, ylabel={},    
  xticklabels=\empty, yticklabels=\empty,
  enlargelimits=false,
]

      \draw[very thick, color=blue] 
          (axis cs:1,5) -- 
          (axis cs:1,4) -- 
          (axis cs:2,2) -- 
          (axis cs:2,3) --
          (axis cs:2,4) --
          (axis cs:3,2) --
          (axis cs:3,3) --
          (axis cs:3,4) --
          (axis cs:4,2) --
          (axis cs:4,3) --
          (axis cs:5,1) --
          (axis cs:5,2) --
          (axis cs:5,3) --
          (axis cs:6,1) --
          (axis cs:6,2) --
          (axis cs:6,3) --
          (axis cs:7,1) --
          (axis cs:7,2) --
          (axis cs:7,3);

      \draw[very thick, color=blue] 
          (axis cs:1,5) -- 
          (axis cs:2,3);

      \node[polynode] at (axis cs:1,5) {};
      \node[polynode] at (axis cs:1,4) {};
      \node[polynode] at (axis cs:2,4) {};
      \node[polynode] at (axis cs:2,3) {};
      \node[polynode] at (axis cs:2,2) {};
      \node[polynode] at (axis cs:3,4) {};
      \node[polynode] at (axis cs:3,3) {};
      \node[polynode] at (axis cs:3,2) {};
      \node[polynode] at (axis cs:4,3) {};
      \node[polynode] at (axis cs:4,2) {};
      \node[polynode] at (axis cs:5,3) {};
      \node[polynode] at (axis cs:5,2) {};
      \node[polynode] at (axis cs:5,1) {};
      \node[polynode] at (axis cs:6,3) {};
      \node[polynode] at (axis cs:6,2) {};
      \node[polynode] at (axis cs:6,1) {};
      \node[polynode] at (axis cs:7,3) {};
      \node[polynode] at (axis cs:7,2) {};
      \node[polynode] at (axis cs:7,1) {};

   \end{axis}

  \end{tikzpicture}
}
\hspace{2cm}
\subfloat[]{
\begin{tikzpicture}[
    scale=0.5,
    polynode/.style={draw=black, circle, fill=black, inner sep=2pt}
    ]
\begin{axis}[
  width=8cm, height=6cm,   
  axis equal image,        
  xmin=0, xmax=8,
  ymin=0, ymax=6,
  grid=both,
  minor tick num=1,
  xlabel={}, ylabel={},    
  xticklabels=\empty, yticklabels=\empty,
  enlargelimits=false,
]

      \draw[very thick, color=blue,-latex] 
          (axis cs:1,5) -- 
          (axis cs:1,4) -- 
          (axis cs:2,2);

      \draw[very thick, color=blue,-latex] 
          (axis cs:5,2) --
          (axis cs:5,3) --
          (axis cs:6,1) --
          (axis cs:6,2) --
          (axis cs:6,3);

      \draw[very thick, color=blue,-latex] 
          (axis cs:3,3) --
          (axis cs:3,2)--
          (axis cs:2,4);

      \draw[very thick, color=blue,-latex] 
          (axis cs:7,3) --
          (axis cs:7,2)--
          (axis cs:7,1);

      \draw[very thick, color=red,-latex] (axis cs:2,2) to[bend right]  (axis cs:5,2); 
      \draw[very thick, color=red,-latex] (axis cs:6,3) to[bend right]  (axis cs:3,3); 
      \draw[very thick, color=red,-latex] (axis cs:2,4) to[bend left]  (axis cs:7,3);

      \node[polynode] at (axis cs:1,5) {};
      \node[polynode] at (axis cs:1,4) {};
      \node[polynode] at (axis cs:2,4) {};
      \node[polynode] at (axis cs:2,3) {};
      \node[polynode] at (axis cs:2,2) {};
      \node[polynode] at (axis cs:3,4) {};
      \node[polynode] at (axis cs:3,3) {};
      \node[polynode] at (axis cs:3,2) {};
      \node[polynode] at (axis cs:4,3) {};
      \node[polynode] at (axis cs:4,2) {};
      \node[polynode] at (axis cs:5,3) {};
      \node[polynode] at (axis cs:5,2) {};
      \node[polynode] at (axis cs:5,1) {};
      \node[polynode] at (axis cs:6,3) {};
      \node[polynode] at (axis cs:6,2) {};
      \node[polynode] at (axis cs:6,1) {};
      \node[polynode] at (axis cs:7,3) {};
      \node[polynode] at (axis cs:7,2) {};
      \node[polynode] at (axis cs:7,1) {};

   \end{axis}

  \end{tikzpicture}
}
\caption{Fiber walks using Markov-basis on left-hand side and SAT-sampler proposals (in red) combined with Markov
moves (in blue) on right-hand side.}\label{fig:sat_vs_markov}
\end{center}
\end{figure}
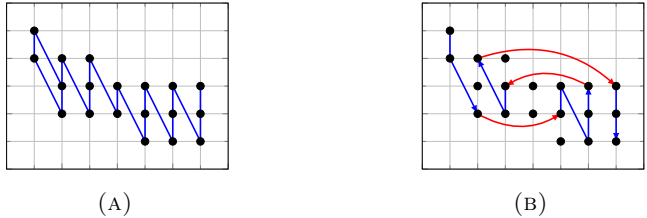

We explore a different approach by encoding fibers as solutions to constraint satisfaction problems
and solve them using SAT and SMT solvers. Despite
SAT’s $\mathrm{N}\mathbb{P}$-completeness, modern conflict-driven clause-learning (CDCL) solvers routinely handle large,
structured instances with
remarkable effectiveness. Representative systems include
CryptoMiniSat5, a Minisat-derived solver extended with native XOR reasoning for cryptographic
constraints \cite{soos2016cryptominisat,sorensson2010minisat}, and more recent high-performance
solver such as Kissat~\cite{kissat}.
Recently, there were advances in \emph{sampling} from SAT instances (e.g.
UniGen3~\cite{unigen3}, CMSGen~\cite{cmsgen}) and from SMT instances
(e.g. SMTSampler~\cite{smtsampler}, GuidedSampler~\cite{guidedsampler}
and MeGASampler~\cite{MeGASampler}), allowing
faster and (almost-)uniform sampling from problem instances. While SMT
samplers claim to be faster in most instances than using bit-blasting
and then SAT samplers, especially the most recent and for our problem
applicable sampler GuidedSampler, our initial experiments seemed to indicate that this might not be
the case for our problem. Performance of these competing approaches depend on the concrete problem
instances. See e.g. \cite{shaw2024csb}, another recent result which gets better results on their
problem instances using bit-blasting and
SAT sampling over SMT sampling. We therefore focus on bit-blasting and applying
SAT samplers. These SAT-based tools have found
applications in software testing, configurable systems, and bug
synthesis~\cite{Dutra2018Efficient,Heradio2022Uniform,Xiang2022Sampling,roy2018bug}.

Our premise is that fibers underlying exact conditional tests can be encoded compactly as Boolean
circuits and then translated to conjunctive normal form (CNF) via a Tseitin-style transformation.
This encoding captures the linear margin constraints, structural zeros, and bound constraints that
define the fiber. Once in CNF, SAT samplers can be used to draw tables from the fiber; enumerating SAT solutions
and applying extraction routines can enumerate tables exactly for small instances; and CDCL machinery can act
as a highly optimized engine for constraint propagation and randomized exploration.
However, integrating SAT-based sampling methods into the fiber-walk framework for exact conditional
inference requires a careful design as SAT samplers typically do not yield perfectly uniform samples
and the transition probabilities between fiber elements are not explicitly known. We address these
issues by only selectively incorporating SAT proposals into hybrid MCMC schemes and still
maintaining the correct stationary distribution after applying Metropolis-Hastings corrections (see
Figure~\ref{fig:sat_vs_markov}). We stress that our goal is not to render Markov bases
obsolete: they remain a valuable and general tool, and our hybrid schemes deliberately retain
local Markov moves. Rather, we target the two bottlenecks discussed above—and the slow-mixing
one in particular—by drawing global SAT proposals without computing a Markov basis beforehand.
Our SAT-viewpoint thereby offers new advantages as it avoids the upfront cost of a global Markov
basis and yields mixing benefits particularly in settings where the combinatorial
structure of the fibers is complex like in the presence of structural zeros or small right hand
sides. Closest in spirit to our work are \emph{dynamic Markov
bases}~\cite{dobra2011dynamicmarkovbases}, which compute moves on the fly, tailored to the current
region of the fiber. We share the central motivation: abandon the precomputation of a full Markov
basis and instead generate proposals adapted to the current table. Dynamic Markov basis compute
moves dynamically by solving a bounded integer linear program that respects the fiber constraints.
Our SAT proposals, by contrast, are drawn as complete solutions of the constraint system and
therefore correspond to entire tables in the fiber, enabling long-range jumps rather than
incremental perturbations. Moreover, structural zeros and bound constraints are encoded natively in
the Boolean circuit, whereas in the dynamic-basis setting they complicate the construction of valid
moves.

This work is organized as follows. After reviewing log-linear models and fiber graphs in
Section~\ref{s:basics}, we provide an efficient encoding of fiber constraints into Boolean circuits
and a Tseitin translation to CNF that preserves sparsity and enables strong unit propagation in CDCL
solvers in Section~\ref{s:sat_sampling}. 
In Section~\ref{s:designing_fiber_walks}, we quantify the extent and structure of sampling
biases in state-of-the-art SAT samplers and develop practical mitigation techniques that improve the
accuracy of conditional $p$-values without sacrificing correctness.
Finally, in Section~\ref{s:benchmarks}, we present empirical results from a comprehensive benchmark of
fiber-walk strategies for contingency tables with and without structural zeros demonstrating that
SAT-based methods can outperform traditional Markov-basis approaches in terms of mixing speed.

\section{Log-linear models and fiber graphs}\label{s:basics}

Let $\Omega=\{\omega_1,\ldots,\omega_d\}$
and let $\cP_\Omega$ the set of all probability mass
functions
$\pi: \Omega\to[0,1]$ on $\Omega$, i.e.
$\sum_{i=1}^d\pi(\omega_i)=1$. To ease notation, we will associate the probability mass
functions on $\Omega$ with vectors $\pi\in[0,1]^d$ having $\|\pi\|_1=1$ by
setting
$\pi_i:=\pi(\omega_i)$.
The log-linear model associated to
$A\in\ZZ^{k\times d}$ is the set of distributions on $\Omega$:
$$\cP_A:=\left\{\pi\in\cP_\Omega:
    (\log\pi_1,\dots,\log\pi_d)\in\mathrm{rowspan}(A)\right\}.$$
A typical question in statistical inference is whether iid samples $x_1,\ldots,x_n\in\Omega$
of an unknown probability mass function
$\overline\pi\in\cP_\Omega$ provide evidence for or against $\overline\pi\in\cP_A$.
Log-linear models constitute a broad and general class of models including, for
instance, hierarchical models. A prominent log-linear model is the
\emph{independence model}~\cite{drton2008}, where two jointly distributed categorical variables should be tested on
statistical independence, as well as the \emph{no-three way interaction model}~\cite[Section~1.2]{drton2008}.
Typically, the frequency counts $u^{\mathrm{obs}}\in\NN^d$ with
$u^{\mathrm{obs}}_i:=|\{j\in[n]: x_j=\omega_i\}|$ for $i\in[d]$ of the observations is the central
object of the analysis, which is presented in many log-linear models as a multi-dimensional \emph{contingency table}.
One can think of a contingency table as a histogram of the observed data.
Statistical significance tests become more involved in degenerated cases,
like if the tables are \emph{incomplete}, that is if certain entries
$S\subset [d]$ called \emph{structural zeros} are impossible and thus $\pi_{i}=0$ for all
$i\in S$. Typically, structural zeros can be included into the model by unit vectors $e_i$, $i\in S$
as rows (see also Section~\ref{s:fibers}).
Now, to test $\overline\pi\in\cP_A$, a first step is to compute a \emph{maximum-likelihood estimator} $\tilde{\pi}$ from $\cP_A$
for the samples $x_1,\dots,x_n\in\Omega$, that is an
element of (the possibly empty set)
$\arg\max_{\pi\in\cP_A}\prod_{i=1}^n\pi(x_i)$.
Let $b:=A\cdot u^{\mathrm{obs}}$ be the \emph{marginals}, then the set of all
frequency counts having the same marginals under $A$ is called
the $b$-\emph{fiber} of $A$: $\fiber{A}{b}=\{u\in\NN^d: A\cdot u=b\}$.
An \emph{extremness-measure} $X:\NN^d\to\RR$
quantifies how extreme a given frequency count $u\in\NN^d$ with
$\|u\|_1=n$ is compared to the frequencies obtained by the maximum
likelihood estimator (see Algorithm~\ref{alg:evaluation} for a popular choice).
Using $X$ and defining
\begin{equation}\label{equ:target_distribution}
    \rho: \fiber{A}{b}\to[0,1],  \rho(v)\sim\frac{1}{v_1!\cdots v_d!}
\end{equation}
and $f: \fiber{A}{b}\to\{0,1\}$ with $f(v)=\mathbf{1}_{X(v)\ge X(u^\textnormal{obs})}$
the value $\mathbb{E}_\rho[f]=\sum_{u\in\fiber{A}{b}}f(u)\rho(u)$ constitutes the \emph{conditional
    $p$-value} that the unknown distribution
$\overline\pi$ yielding the frequency counts
$u^\text{obs}$ is from $\cP_A$.
Clearly, the normalizing constant of $\rho$ together with the size of $\fiber{A}{b}$ make it hard to
compute $\mathbb{E}_p[f]$ directly and thus sampling from $\fiber{A}{b}$ is required to approximate
it.

A classic way to approximate conditional $p$-values is to construct a Markov chain that
converges against $\rho$ on $\fiber{A}{b}$ using the \emph{Metropolis-Hastings}
methodology~\cite{metropolis1953,hastings1970}. Generally, given a graph $G$
on $\fiber{A}{b}$ and a random walk $\cW$, the \emph{Metropolis-Hastings walk}
    $\metropolis{\cW}{\rho}$ is the random walk on $G$ defined by
    \begin{equation*}
        \metropolis{\cW}{\rho}(u,v):=\begin{cases}
            \cW(u,v)\cdot\min\left\{1,\frac{\rho(v)\cW(v,u)}{\rho(u)\cW(u,v)}\right\}, & \textnormal{ if
            }u\neq v                                                                                     \\
            1-\sum_{w\in\fiber{A}{b}\setminus\{u\}}
            \cW(u,w)\cdot\min\left\{1,\frac{\rho(w)\cW(w,u)}{\rho(u)\cW(u,w)}\right\}, & \textnormal{ if
            }u=v                                                                                         \\
        \end{cases}.
    \end{equation*}

Many properties of the random walk $\mathcal{W}$ hold also for $\metropolis{\cW}{\rho}$.
Particularly, if $\cW$ is irreducible, aperiodic,
and if $\cW(u,v)>0$ if and only if $\cW(v,u)>0$ for all $u,v\in\fiber{A}{b}$,
then $\metropolis{\cW}{\rho}$ is as well irreducible, aperiodic and
its stationary distribution is $\rho$~\cite[Lemma~1.1]{diaconis1998-mcmc}.
Moreover, running a Metropolis-Hastings walk in practice is simple:
Suppose the random
walk is at node $u\in\fiber{A}{b}$, then we sample $v\in\fiber{A}{b}$ according to
$\cW(u,\cdot)$ and compute
$$\reject{\cW}{u}{v}:=\min\left\{1,\frac{\cW(v,u)}{\cW(u,v)}\cdot\prod_{i=1}^d\frac{u_i!}{v_i!}\right\}$$
where we used $\rho$ from Equation~\ref{equ:target_distribution}.
In a
second step, we walk to $v$ with probability $p$ and we stay at $u$
with probability $1-p$ (see Algorithm~\ref{alg:mh} in Section~\ref{app:mh}). Clearly, if $\cW$ is symmetric, $\reject{\cW}{u}{v}$
simplifies further.
In order to construct a random walk $\cW$ on $\fiber{A}{b}$ feasible for Metropolis-Hastings, we not
only need irreducibility, i.e. all elements in the fiber can be reached, but also control over the
transition probabilities to compute $\reject{\cW}{u}{v}$.
One way to construct a random walk is by using tools of commutative algebra
as proposed in the seminal paper~\cite{Diaconis1998a}, where one has to
compute a \emph{Markov basis} $\cM\subset\ZZ^d$ for $A$, which is a set of vectors whose corresponding binomials in the
polynomial ring constitute a generator set of the toric ideal of $A$. Geometrically, a Markov basis
allows to connect any two elements in any fiber of a given matrix $A$. We denote the
corresponding transition matrix with $\cW_{\text{MB}}$. 
The obtained graph on the fiber is called \emph{fiber
graph}. Typically, scenarios with finite fibers are considered, where $A$ can be
chosen to have non-negative entries only.

\section{Fibers as SAT-Instances}\label{s:sat_sampling}
In this section, we present our approach to encode $\fiber{A}{b}$ for given
$A \in \mathbb{N}^{k \times d}$ and $b \in \mathbb{Z}^k$ as a SAT-instance.
By definition, $\fiber{A}{b}$ is the set of all solutions $u \in \NN^d$ to
the $k$ many equations $A_iu=b_i$ and hence it can be seen as an instance of a constraint
satisfaction problem over $\NN$. 
SAT-solvers are designed to solve constraint satisfaction problems over boolean algebras and hence
their solution space is in $\{0,1\}^{q}$ for some $q\in \NN$. Equivalently, the solution space can be
$[2^l-1]^d$ using $l \in \NN$ bits to represent each entry. So we can map fiber elements one-to-one by representing $u \in \NN^d$ as a $d$-ary vector of $l$-bit numbers for some $l \in \NN$. Note that $l$ needs to be at most $\lceil\log_2(\max(b))\rceil$.
Thus, if we would only consider solutions $u$ whose
entries have at most $l$-many bits, i.e. $u\in[2^l-1]^d$, then generating elements from
$\fiber{A}{b} \cap ([2^l-1])^d$ can be reduced to the satisfiability problem of boolean formulas in
conjunctive normal form (CNF). The equations $Au=b$ can be reformulated as a series of additions, multiplications and
equality checks where each be implemented by a boolean circuit
$C$ on variables $u_1, \ldots, u_d$ such that $C(u_1, \ldots, u_d) = 1$ if and only if $(u_1,
\ldots, u_d)$ is a solution to $Ax=b$. Clearly, the size $C$ is in $\textrm{poly}(k,d,l)$.

Next, we use the classic Tseitin encoding~\cite{schoning2013satisfiability}, which allows to
transform $C$ to a boolean formula $F$ in conjunctive normal form (CNF) which is
satisfiable if and only if $C$ is satisfiable. Moreover, $C(u_1, \ldots, u_d) = 1$ if and only if
there exists exactly one $y_1, \ldots, y_m \in \{0, 1\}^m$ such that $F(u_1, \ldots, u_d, y_1,
\ldots, y_m) = 1$.

Recall that a \emph{gate} is a finite number boolean operation with an output and some inputs, e.g.
output of another gate or input variables. The Tseitin transformation replaces every gate $G$ in $C$
by a CNF formula of size at most linear in the number of inputs of $G$ such that the formula encodes
a correct evaluation of $G$.
For instance, if $G = a \land b$, we encode $G$ by the formula $z \leftrightarrow (a \land b) \equiv (\neg z
\lor a) \land (\neg z \lor b) \land (z \lor \neg a \lor \neg b)$. Observe that for all satisfying
assignments of this formula, given $a, b \in \{0, 1\}$, $z = a \land b$. If we apply this to every
gate in $C$ we get a conjunction of $m \in O(|C|)$ many clauses with at most $m$ many helper
variables encoding the evaluation of $C$. Adding the single clause $(z)$ encoding the output of $C$
enforces all satisfying solutions of $F$ to be satisfying solutions of $C$. Notably, for every
assignment to $(u_1, \ldots, u_d)$ there is only one correct evaluation for $C$ and therefore the
corresponding $y_1, \ldots, y_m$ are unique.
Thus, these transformations yield a bijection between
$\fiber{A}{b}\cap[2^l-1]^d$ and the satisfying solutions to $F$.
We can tighten $l$ even more: it is easy to see that for $l = \lceil \textrm{log}_2(\max_{i,j,
A_{i,j}>0} \frac{b_{i}}{A_{i,j}}) \rceil$ we have $\fiber{A}{b} = \fiber{A}{b} \cap ([2^l-1])^d$.
Particularly, if $A$ has only entries in $\{0,1\}$, like its the case for independence tests of
contingency tables, we need at most as many
bits as the largest entry of $b$. In other words, we can embed $\fiber{A}{b} \cap ([2^l-1])^d$ into the hypercube 
$\{0,1\}^{l\cdot d + |C|}$. This allows us to use SAT-based techniques
to generate fiber elements. In Appendix~\ref{app:sat_encoding}, we provide a walk-through
of the encoding and the resulting CNF formula for a concrete fiber.

Moving from generation of solutions to \emph{uniform} sampling of solutions is more involved.
Ideally, we would like to know the transition probabilities
$\cW_{\mathrm{SAT}}(u, v)$ for two solutions explicitly, but this remains computationally
intractable. Thus, we resort to SAT-samplers that are designed to sample uniformly or almost
uniformly. 
However, it is not easy to obtain uniform sampling by just randomizing variable order or
branching decisions of a CDCL solver. Also adding heuristics, like more
preprocessing and clause learning, prohibits uniform sampling even more. Nevertheless, there exist
SAT-solvers that can generate uniform or almost uniform random solutions to a given CNF.
One is UniGen3~\cite{unigen3}, which uses universal hashing to randomly select areas of the solution
space and generates one solution in this area. Another is CMSGen \cite{cmsgen}, a Cryptominisat5
based SAT-sampler, which is built to fool the statistical test Barbarik \cite{pote2022scalable,barbarik22} for
uniform sampling. We found CMSGen to be faster than UniGen3 while still being somehow uniform.
Let $p_{\mathrm{SAT}}(u)$ be the probability that a SAT-sampler draws the solution $u$ from $\fiber{A}{b}$, then
almost uniform samplers like UniGen3 satisfy $\frac{1-\epsilon}{|\fiber{A}{b}|} \leq
p_{\mathrm{SAT}}(u) \leq
\frac{1+\epsilon}{|\fiber{A}{b}|}$ for a controllable hyperparameter $\epsilon > 0$.
Therefore, the transition probabilities are all
positive giving a complete graph on $\fiber{A}{b}$.
On the other hand, CMSGen~\cite{cmsgen} is designed to be not too far from a uniform sampler, i.e.
the statistical test Barbarik did not conclude that the total variance distance to the uniform is
more than $\eta$, that is, $\sum_{u \in \fiber{A}{b}} |p_{\mathrm{SAT}}(u) -
\frac{1}{|\fiber{A}{b}|}|
    \leq \eta$ is satisfied. 
Particularly, UniGen3 allows to control per-solution bounds while CMSGen only controls the total
distance on the fiber.
\cite{cmsgen} report that the Barbarik test could not distinguish between Unigen3 and CMSGen with
$\eta = 1.8$.
All SAT-samplers discussed in our work generate solutions that are statistically independent of previous sampled solutions.

\section{Designing SAT-based fiber random walks}\label{s:designing_fiber_walks}

For small problem instances, our SAT-encoding can be used to quickly enumerate the full
fiber and to compute the $p$-value exactly.
For larger instances, an integration of SAT-methods into the Metropolis-Hastings framework is
needed, which requires a careful design when convergence to a target distribution is desired.
Particularly, this requires to control the individual transition
probabilities $\cW_{\mathrm{SAT}}(u,v)$ between two fiber elements $u,v\in\fiber{A}{b}$. In general,
these transition probabilities are unknown for SAT-samplers. Thus, we use SAT-samplers designed to
approximate the uniform distribution. Naively, one could solely use a SAT-sampler to sample from the fiber
$\fiber{A}{b}$ directly. 

Clearly, when the SAT-sampler is \emph{exactly} uniform over $\fiber{A}{b}$, i.e.\
$\cW_{\mathrm{SAT}}(u,v) = 1/\lvert\fiber{A}{b}\rvert$ for all $u,v\in\fiber{A}{b}$, then
$\cW_{\mathrm{SAT}}(v,u)=\cW_{\mathrm{SAT}}(u,v)$ and the acceptance probability simplifies to
$$
  \reject{\cW_{\mathrm{SAT}}}{u}{v} = \min\!\left(1,\, \prod_{i=1}^d\frac{u_i!}{v_i!}\right) =
\min\!\left(1,\, \frac{\rho(v)}{\rho(u)}\right).$$
Thus, thus global sampling combined with any local walk using a fixed set of moves $\cM$ yields a
Markov chain that converges to the correct target distribution $\rho$ on $\fiber{A}{b}$,
\emph{independent}
of the choice of $\cM$.

\begin{figure}[!ht]
\begin{center}
    \includegraphics[width=\textwidth]{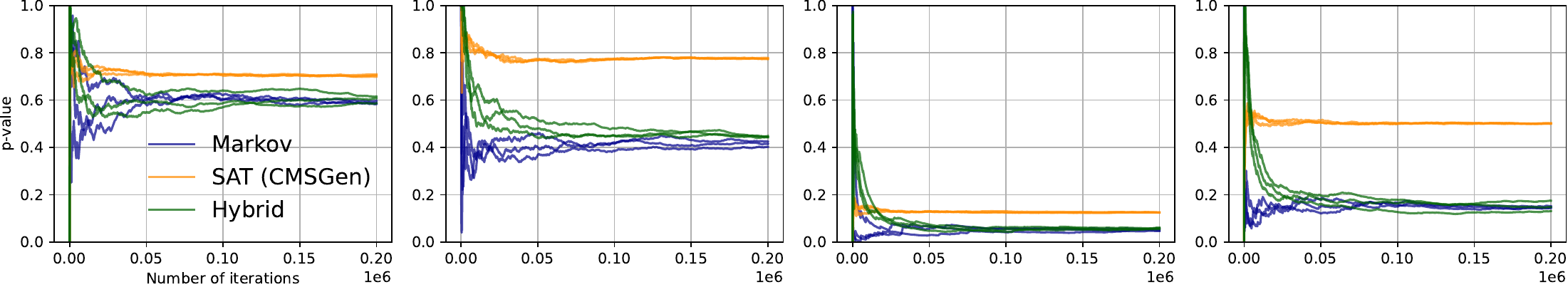}
    \caption{Convergence to $p$-values for independence testing in sparse $10\times 10$
    contingency tables, where SAT-only methods show strong structural biases prohibiting direct use for
approximating $p$-values. In a hybrid combination with Markov basis, chains converge to correct $p$-value.}\label{fig:sampling_bias}
\end{center}
\end{figure}

In our experiments, however, we found that almost uniform sampling
computationally is too expensive already for small problem instances. Although not-too-far-from uniform
samplers like CMSGen allow to process large problem instances, their
structural bias hinders a reliable $p$-value approximation already in small problem instances as
shown in Figure~\ref{fig:sampling_bias}. Consequently, to reduce structural bias but keep the good
mixing properties, we can only use steps from the SAT-samplers selectively.
To use them effectively in combination with existing methods, we use them to ensure irreducibility
of the underlying Markov chain. 
Table~\ref{tab:convergence} summarises the convergence guarantees for both schemes across all combinations of move set and SAT-sampler type.

\subsection{$\parallelAlg$}

First, we introduce a parallel scheme $\parallelAlg$ that initializes $k$ independent walks by
sampling $k$ starting points from $\fiber{A}{b}$ using the SAT-sampler and then evolves each walk
using local moves from $\cM$. The walks proceed sequentially: every $n$ steps, one of the $k$ walks
is selected uniformly at random and advanced for $n$ further steps. Switching walks after a fixed
block of $n$ steps keeps all $k$ chains progressing in tandem rather than running a single one to
completion, so the pooled sample reflects the coverage of the whole ensemble throughout the run:
diverse SAT-initialized starting points—and, when $\cM$ is not a Markov basis, distinct connected
components—are explored in parallel, which speeds up convergence. The block length $n>1$ amortizes the cost of switching and lets each
selected walk take enough local steps to decorrelate from its previous visit before being paused.

$\parallelAlg$ can be understood as a multi-chain variant of the classical Monte Carlo
significance tests of Besag and Clifford~\cite{besag1989generalized,besag1991sequential}.
Their \emph{parallel} construction also runs several dependent chains and pools the resulting
states to estimate a $p$-value. In our scheme the SAT-sampler supplies the global moves that make the ensemble of chains explore the whole
fiber, while the local moves from $\cM$ provide cheap within-component mixing. Viewed this way, the
SAT-steps are the global exploration mechanism of the ensemble and the role of the block length $n$ and
the number of chains $k$ is to trade off how much of the budget is spent on global re-initialization
versus local decorrelation. 

When $\cM$ is a Markov basis, each chain is individually irreducible, so the
SAT-initialization only accelerates convergence. When $\cM$ is not a Markov basis the SAT-steps are what
restores irreducibility across the connected components of $\fibergraph{A}{b}{\cM}$, exactly as the global
proposals do in the Besag--Clifford setting. 
Convergence to $\rho$ on the full
fiber then depends on whether the SAT-sampler places initial points across all components.

\subsection{$\hybridAlg$}\label{s:hybrid}

Our next scheme $\hybridAlg$ is a single-chain hybrid that interleaves local moves from $\cM$ with
global SAT proposals with a fixed ratio: every $n$-th step proposes a
new table via the SAT-sampler, while all other steps use $\cM$.
Clearly, if $\cW_{\mathrm{SAT}}$ is exactly uniform, the SAT proposals are accepted with probability $\reject{\cW_{\mathrm{SAT}}}{u}{v}$ as described above, giving a chain that converges to $\rho$ regardless of whether $\cM$ is a Markov basis, since the SAT steps alone ensure irreducibility on $\fiber{A}{b}$.

For approximately uniform SAT-samplers, the ratio $\cW_{\mathrm{SAT}}(v,u)/\cW_{\mathrm{SAT}}(u,v)$
in $\reject{\cW_{\mathrm{SAT}}}{u}{v}$ is unknown.
Thus, when applying the Metropolis-Hastings correction, we approximate it by $1$ and accept the SAT
proposal with probability same probability then from the moves set, namely $\min\{1, \rho(v)/\rho(u)\}$.
This breaks
detailed balance for the SAT component: the resulting Markov chain has a stationary distribution
$\rho_n$ that is biased away from the target $\rho$.
To analyse the bias, we use the Dobrushin coefficient~\cite{dobrushin_coef} of the local walk induced by a set of moves to
quantify how strongly the local moves contract towards $\rho$ between SAT-steps. 
For any transition matrix $\cW$, we define
$$
c(\cW):=\max_{u,v\in\fiber{A}{b}}\bigl\lVert \cW(u,\cdot)-\cW(v,\cdot)\bigr\rVert_{\mathrm{TV}}\in[0,1].
$$
If $\cW$ comes from a set of moves $\cM$ so that for all $u,v\in\fiber{A}{b}$ there exists a path of
length at most $n$, then $c(\cW^n)<1$. Thus, if $\cM$ is a Markov basis
$c(\cW^n)\to 0$ for $n\to\infty$. Vice versa, if $c(\cW^n)<1$ for some $n$, then $\cM$ gives a
connected graph on $\fiber{A}{b}$.

\begin{thm}\label{thm:hybrid_bias}
Let the SAT-sampler be $\epsilon$-uniform
and let $\cM$ be a set of moves and $n\in\NN$ such that $c(\cW^{n-1})<1$.
Then $\hybridAlg$ has a unique stationary distribution $\rho_n$
which satisfies
\[
  \bigl\lVert \rho_n-\rho\bigr\rVert_{\mathrm{TV}} \;\leq\;\frac{2\epsilon}{(1-\epsilon)\,\bigl(1-c(\cW^{n-1})\bigr)}.
\]
\end{thm}
\begin{proof}
See Appendix~\ref{app:bias}.
\end{proof}

Clearly, if the SAT-sampler is exactly uniform, i.e. $\epsilon=0$, then $\rho_n=\rho$.
Moreover, the bias is always of order $\epsilon$ and is governed by
$c(\cW^{n-1})$, which decreases monotonically in $n$ towards $0$. 

\begin{table}[ht]
\centering
\begin{tabular}{llcc}
\toprule
\multirow{2}{*}{$\cM$} & & \multicolumn{2}{c}{SAT-sampler} \\
\cmidrule(l){3-4}
& & exact uniform & approx.\ uniform \\
\midrule
\multirow{2}{*}{Markov basis}
  & $\parallelAlg$ & $\checkmark$ & $\checkmark$ \\
  & $\hybridAlg$   & $\checkmark$ & $\checkmark{}^\dagger$ \\
\midrule
\multirow{2}{*}{no Markov basis}
  & $\parallelAlg$ & $(\checkmark)$ & $(\checkmark)$ \\
  & $\hybridAlg$   & $\checkmark$   & $\times$ \\
\bottomrule
\end{tabular}
\caption{Convergence guarantees for $\parallelAlg$ and $\hybridAlg$ by move set and SAT-sampler type.
$\checkmark$: converges to $\rho$;
$(\checkmark)$: converges to $\rho$ restricted to each connected component of $\fibergraph{A}{b}{\cM}$;
$\times$: not guaranteed.
${}^\dagger$ Requires keeping the fraction of SAT-steps sufficiently small.}
\label{tab:convergence}
\end{table}

\section{Benchmark study}\label{s:benchmarks}

To measure effectiveness and efficiency of a sampling method, counting the number of samples generated by a
sampler as done in prior work is not insightful when it comes to approximating $p$-values.
For instance, a rapidly
mixing Markov chain needs far fewer samples than a slowly mixing one.
Instead, we empirically
measure the convergence speed of the generated sequence $\frac{1}{j} \sum_{i=1}^j f(u_i)$ to the
conditional $p$-value $\mathbb{E}_\rho[f]$ of Fishers test.
Convergence speed depends on
many factors, from dimensionality, sample size, to the actual initial observation. Thus
we measure convergence for different constraint matrices, sample sizes, and initial observations.
Different than existing work, we do not focus on the number of samples generated but on the number
of steps required to reach a certain accuracy in terms of approximating the conditional $p$-value
which is more meaningful in practice. Our evaluation scheme is shown in Algorithm~\ref{alg:evaluation} in detail. For a given constraint
matrix $A$, we first construct initial samples $\{\hat u^1,\ldots, \hat u^T\}$, where $T=100$ in our
experiments. Then we run Fishers test for each $\hat u^i$ using the Metropolis-Hastings walk for a
predefined number of steps and measure the speed of their convergence to their
limit. At the back end, Metropolis-Hastings uses a vanilla Markov basis sampler and the proposed
SAT-sampler. In our experiments, the following model classes are considered:

\begin{itemize}
    \item{$\mathrm{I}_{d_1\times \cdots\times d_k}$: The $d_1\times d_2\times\cdots\times d_k$-independence model.}
    \item{$\mathrm{QI}_{d_1\times \cdots\times d_k}(S)$: $\mathrm{I}_{d_1\times \cdots\times d_k}$ with structural zeros
          $S\subseteq[d_1]\times\cdots\times[d_k]$.}
    \item{$\mathrm{N3F}_{d}$: The no-3-factor interaction model on $d\times d\times d$
          tables.}
\end{itemize}
We explain these models in detail in Section~\ref{s:fibers}.
Note that the matrix of the sufficient statistics have their entries all in $\{0, 1\}$.
In the reminder of this section, we state implementation details of Algorithm~\ref{alg:evaluation}
for the different models used. We highlight that currently, we report the number of
steps used by the sampler only, not the processing time required as this depends heavily on the system
running the evaluation. Clearly, one benefit of samplers running with Markov bases are the low
system requirements compared with SAT-samplers. Also, the samples/second of SAT-samplers depend on the problem size (number of boolean variables, number of clauses) and structure. On the other hand, computing a Markov basis beforehand is expensive and
requires very high system requirements. Thus, a comparison in terms of processing time between Markov basis
samplers and SAT samplers cannot be made fair.
For $\mathrm{QI}_{5\times 5}$, we were able to compute the exact conditional $p$-value by enumerating the
whole fiber using our SAT-encoding. This allows us to measure the actual convergence speed to the
exact $p$-value.

\begin{algorithm}
    \caption{Evaluation scheme}
    \SetKwInOut{Input}{Input}
    \SetKwInOut{Output}{Output}
    \Input{$A\in\RR^{k\times d}$, number of runs $T$, sample size $n$, number of iterations $N$}
    \Output{Sequences $p_1,\ldots,p_T\in[0,1]^N$}
    Generate initial values $\hat u^1,\ldots,\hat u^T\in\mathbb{N}_0^d$ with $\|\hat u^i\|_1=n$\\
    \For{\(i \in [T]\)}{
        \uIf{with structural zeros}{
            Compute $S\subseteq\{j\in[d]: (\hat u^i)_j=0\}$ such that MLE exists for $\mathcal{P}_{A, S}$
        }
        \Else{
            $S=\emptyset$
        }
        Compute MLE $\tilde\pi\in\mathcal{P}_{A, S}$ for $\hat u^i$ under $\mathcal{P}_{A, S}$\\
        Define $X:\RR^d\to\RR_{\ge 0}, X(u):=\sum_{i=1}^d\frac{(\frac{u_i}{n}-\tilde\pi_i)^2}{\tilde\pi_i}$\\
        Run Metropolis-Hastings walk $(\hat u^i, X, N)$ and write result to $p_i\in\RR^N$
    }
    \Return{$p_1,\ldots,p_T$}
    \label{alg:evaluation}
\end{algorithm}

To be agnostic to the model family and structural zeros, we implemented a generic gradient-descent based MLE
computation for log-linear models with constraint matrix $A\in\NN^{k\times d}$
and structural zeros $S\subseteq[d]$ and initial observation $u^{\mathrm{obs}}\in\NN^d$.
For a parameter $\theta\in\mathbb{R}^{k}$ we define probabilities on $\overline{S}$ by the softmax of the linear scores
$\pi_\theta(i)\sim \exp(A^\top\theta)_i$ for $i\in\overline{S}$ and $\pi_\theta(i)=0$ for $i\in S$.
We want to compute $\hat\theta$ that maximizes the log-likelihood
$$\ell(\theta)=\sum_{i\in\overline{S}} u^{\mathrm{obs}}_i\cdot\log \pi_\theta(i)$$
yielding $\hat u(\theta):=\|u^{\mathrm{obs}}\|_1\cdot \pi_\theta$ as MLE estimate.
We optimize
$\ell$ with L-BFGS on $\theta$ and monitor the margin discrepancy $\|A(u^{\mathrm{obs}}-\hat
u(\theta)\|_2$ during optimization and reduce the learning rate if the margin discrepancy does not
decrease for a number of iterations.

\subsection{Sufficient statistics matrices}\label{s:fibers}

\subsubsection{Independence models}\label{s:indep}

Given a $d_1\times\cdots\times d_k$ contingency table $u$, the
sufficient statistics matrices $A_{d_1,\ldots, d_k}$ of $I_{d_1\times\cdots\times d_k}$ compute the
$d_1+d_2+\ldots+d_k$ many
$1$-marginals of $u$. In our experiments, we consider models on two- and three-way contingency
tables, which are well studied in both, statistics and combinatorics.
For instance, for the two-way independence model, $A_{d_1,d_2}$ corresponds to the node-edge incidence matrix of the complete bipartite graph
$K_{d_1,d_2}$ and the elements of a Markov basis correspond to $4$-cycles in $K_{d_1,d_2}$. The
elements of this Markov basis are often called \emph{basic moves} in the literature. In our
experiments, we use the Markov basis consisting of basic moves for the Markov basis sampler.

If all $1$-margins are strictly larger than zero, the MLE and a closed-form solutions exists,
see for instance~\cite{existence_mle_quasi_independence}.
For $I_{d_1\times d_2}$, the $1$-marginals are exactly the row sums $r_1,\ldots,r_{d_1}$ and column
sums $c_1,\ldots,c_{d_2}$ and the MLE is $\tilde\pi_{i,j}:=\frac{r_i\cdot c_j}{n\cdot n}$ where
$n=\|u\|_1$ is the sample size (see for instance~\cite[Example~2.1.2]{drton2008}). Thus, when
creating the initial tables, we ensure that the margins are positive.
When creating the initial
two-way tables $\hat u^i$, we make sure that we have varying distance to the expected frequencies
$n\cdot\tilde\pi$. Particularly, we set
$\hat u^i=\lambda\cdot u_{\mathrm{indep}}+ (1-\lambda)\cdot u_{\mathrm{dep}}$
where $\lambda\in[0,1]$ and where $u_{\mathrm{indep}}$ and $u_{\mathrm{dep}}$ are randomly generated
tables whose margins are independent and dependent respectively.

\subsubsection{Two-way quasi-independence model}

Here, the sufficient statistics is the same as for $I_{d_1\times d_2}$, but additionally,
a set $S\subset [d_1]\times[d_2]$ of structural zeros is given and a walk on
$$\fiber{A}{b}\cap\{u\in\NN^d:u_s=0 ~\forall s \in S\}$$
needs to be performed.
Generally, the MLE does not exists for quasi-independence models making
statistical analysis challenging~\cite{fienberg_incomplete_tables}.
More about the uniqueness and existence of MLE in quasi-independence models can be found
in~\cite{existence_mle_quasi_independence}. In~\cite{mle_quasi_independence},
an explicit formula for the MLE $\tilde\theta$ given an observation $u^{\mathrm{obs}}$ in cases the MLE exists
and is rational (see~\cite[Theorem~5.4]{mle_quasi_independence}) was given. More specifically, their formula
is valid for cases where the bipartite graph $K_{d_1,d_2}\setminus S$, that is the complete
bipartite graph where the edges $(i,j)\in S$ have been removed, is \emph{doubly chordal bipartite}.
To generate the initial values in a way the MLE exists, we create an initial table $u$ exactly as in
Section~\ref{s:indep} for the independence model. We then define the set of structural zeros
initially to $S=\{(i,j): \hat u_{i,j}=0\}$ and checking
whether $K_{d_1,d_2}\setminus S$ is doubly chordal bipartite. If not, we add a random edge to a
cycle having less than two chords and extend $S$ in that way until the graph is doubly chordal
bipartite. Note that the basic moves are not guaranteed to connect all elements in
$\fiber{A}{b}\cap\{u\in\NN^d: \forall s \in S~u_s=0\}$. We refer to~\cite{quasi_indep_markov_basis}
and~\cite{subsum_problems} for a deeper analysis on when this is the case. Thus, we have to use the
\emph{Graver basis}~\cite[Section~3]{Loera2013} in
general. Geometrically, the elements of the Graver basis corresponds to the set of all cycles in
$K_{d_1,d_2}\setminus S$ and thus the Graver basis grows exponentially with the dimensions of the
tables.

\subsubsection{No-three way interaction model}

Mathematical descriptions for Markov bases of no-three way interaction models only exist in
special cases, like for $3\times d\times d$ (see~\cite{no3f_3kk}). In our experiments, we used
precomputed Markov bases from \url{https://markov-bases.de} for $3\times 3\times 3$ and $4\times
    4\times 4$ consisting of $81$ and $148.968$ elements respectively.
For a given sample size $n$, we sample $n$ many indices from $[d]\times[d]\times[d]$ uniformly to obtain
a random three-way table $v^i$ for all $i \in [T]$. We than used the SAT-sampler to get an almost uniform element $\hat u^i$ from the fiber
with margins $A\cdot v^i$. 

\subsection{Results}

\begin{figure}[!ht]
\begin{center}
    \subfloat[$\mathrm{N3F}_{4}$]{\label{fig:n3f_4}
        \includegraphics[height=0.18\textheight]{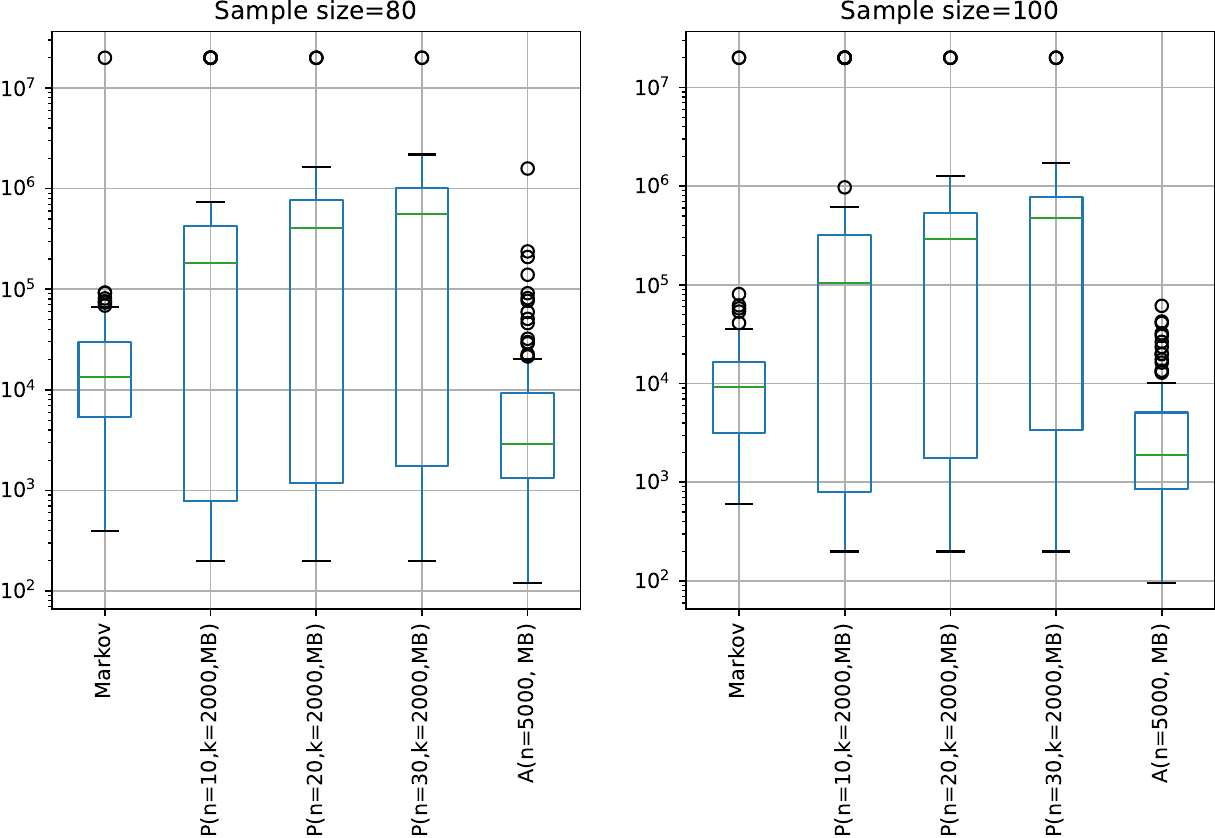}
    }
    \subfloat[$\mathrm{I}_{5\times 2\times 2}$]{\label{fig:522}
        \includegraphics[height=0.18\textheight]{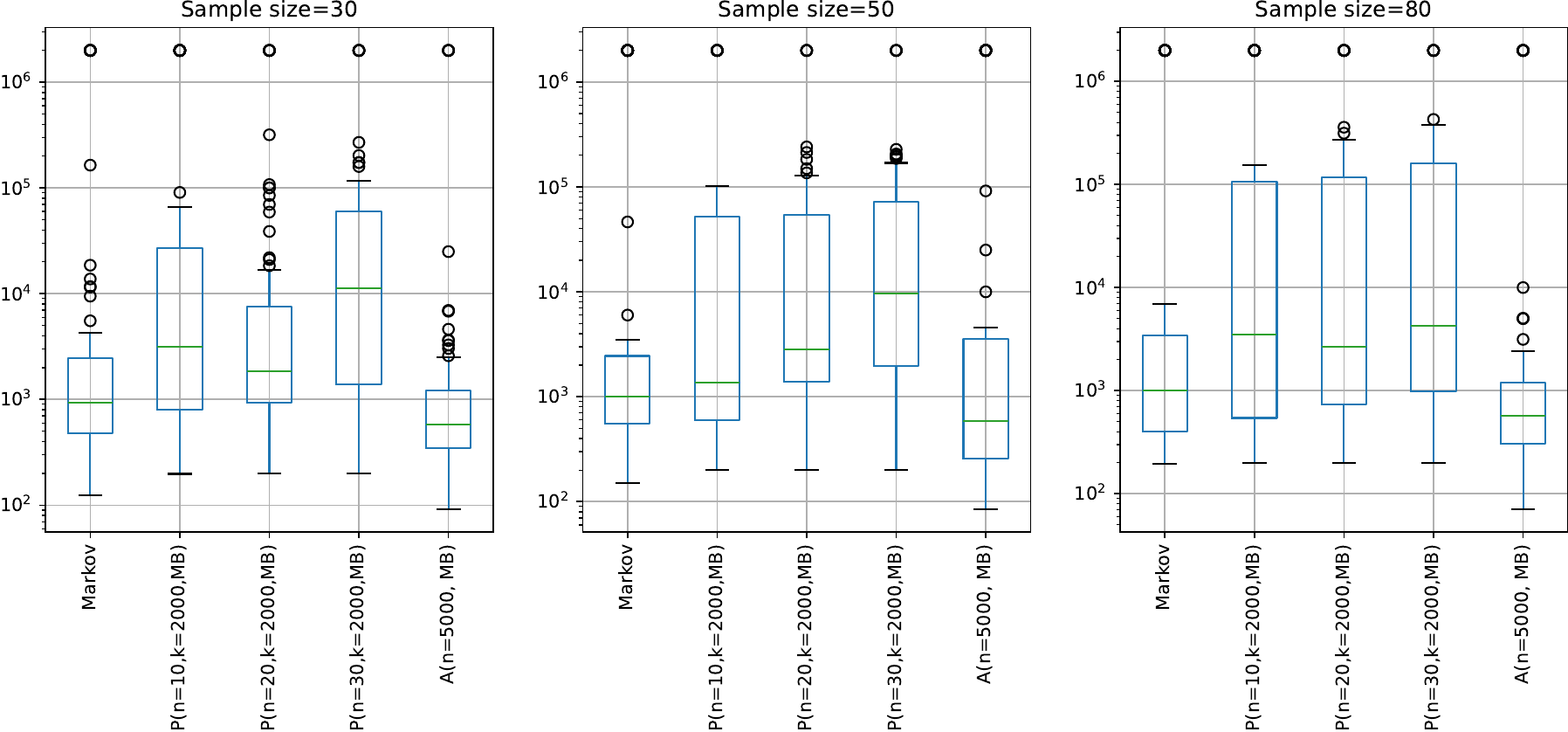}
    }

    \subfloat[$\mathrm{QI}_{5\times 5}$]{
        \includegraphics[height=0.18\textheight]{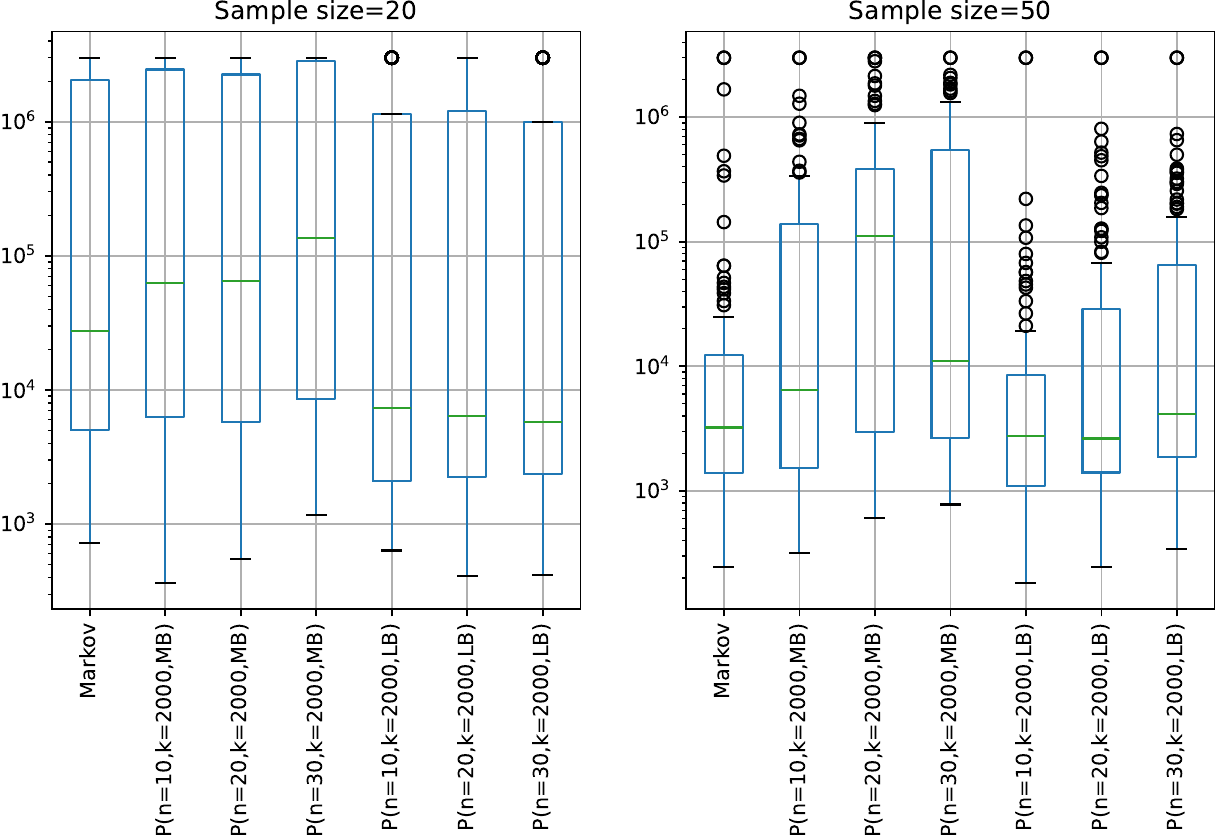}
    }
    \subfloat[$\mathrm{QI}_{10\times 10}$]{\label{fig:sat10x10_wth_bm}
        \includegraphics[height=0.18\textheight]{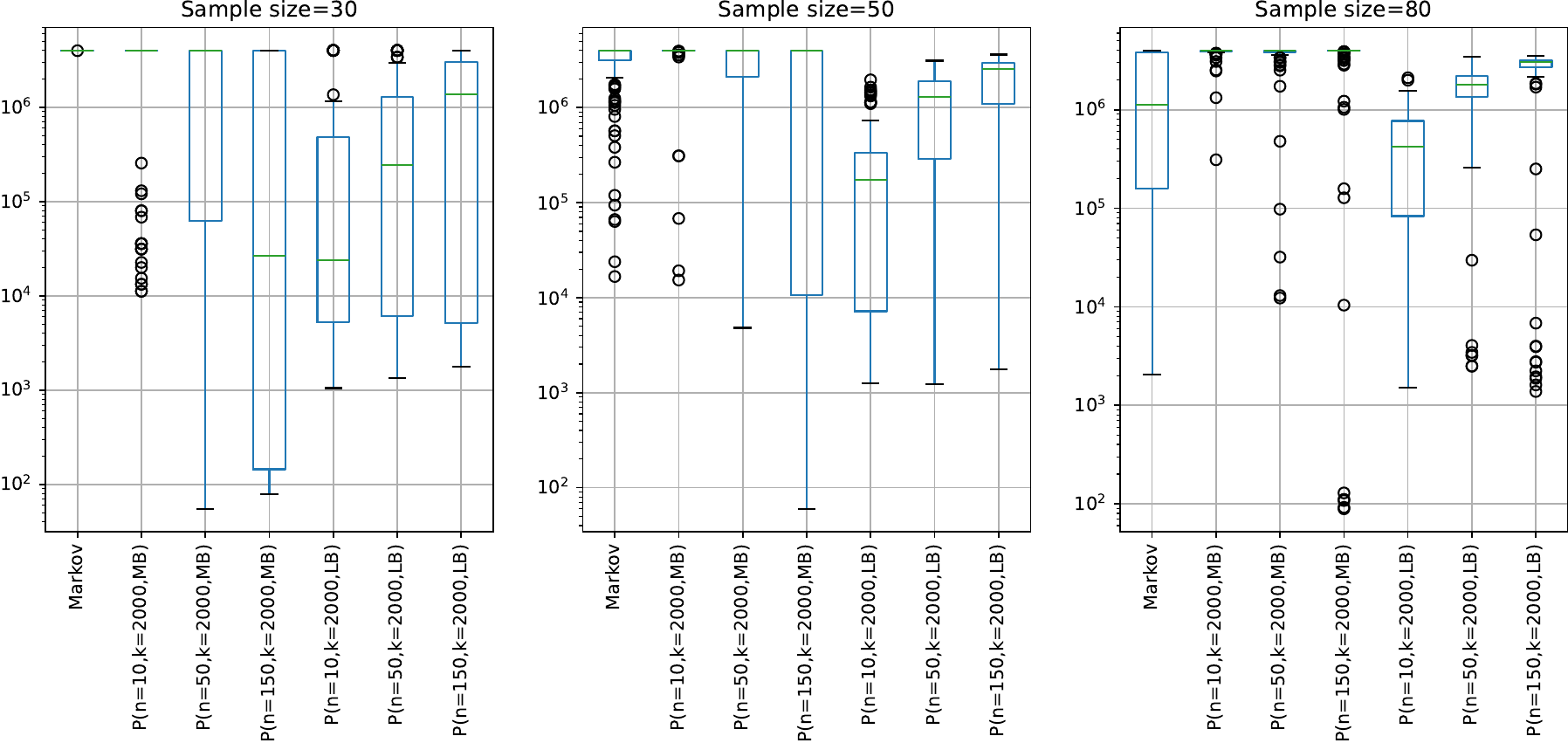}
    }
    \caption{Selected results where SAT-based random walks outperform Markov bases based random
    walks.}
\end{center}
\end{figure}

We observe that the proposed SAT-based random walks outperform Markov basis based random walks in
many scenarios, especially in involved examples with small sample sizes or a large number of
structural zeros. Selected results are shown in Figure~\ref{fig:n3f_4}--\ref{fig:sat10x10_wth_bm}.
There, we show the number of steps required to be within $0.005$ of the final limit. For smaller
instances, like $\mathrm{QI}_{5\times 5}$, we compute convergence against the true $p$-value by enumerating the whole fiber
using our SAT-encoding. Generally, we observe that our SAT-based methods converge faster than Markov
basis based methods when tables are sparse, that is the sample size of the tables is small or
structural zeros are present. This effect seems to be alleviated for larger sample sizes as the
fibers become larger and typically better connected by local moves. In scenarios like
$\mathrm{I}_{k\times k}$, not much difference in convergence can be reported.  Notably, for larger
problem instances, $\hybridAlg$ outperforms $\parallelAlg$ as shown in Figure~\ref{fig:n3f_4},
Figure~\ref{fig:522}, and Figure~\ref{fig:ik22}.
We observed that for quasi-independence models, there is no need to use a full Markov
basis, like the Graver basis, to ensure convergence to the correct $p$-value. Surprisingly, we
observe a
faster convergence when using only the basic moves together with SAT-steps as shown in
Figure~\ref{fig:qi5x5_no_markov_basis}. 
All results of our experiments are shown in the appendix in Section~\ref{a:results}.
\begin{figure}[!ht]
\begin{center}
\subfloat[]{
    \includegraphics[height=0.18\textheight]{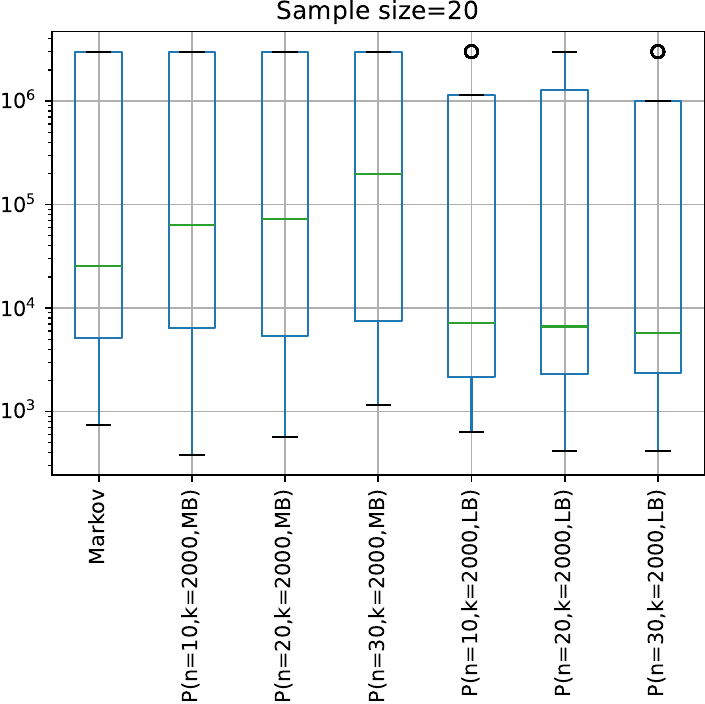}
}
\subfloat[]{
    \includegraphics[width=0.5\textwidth]{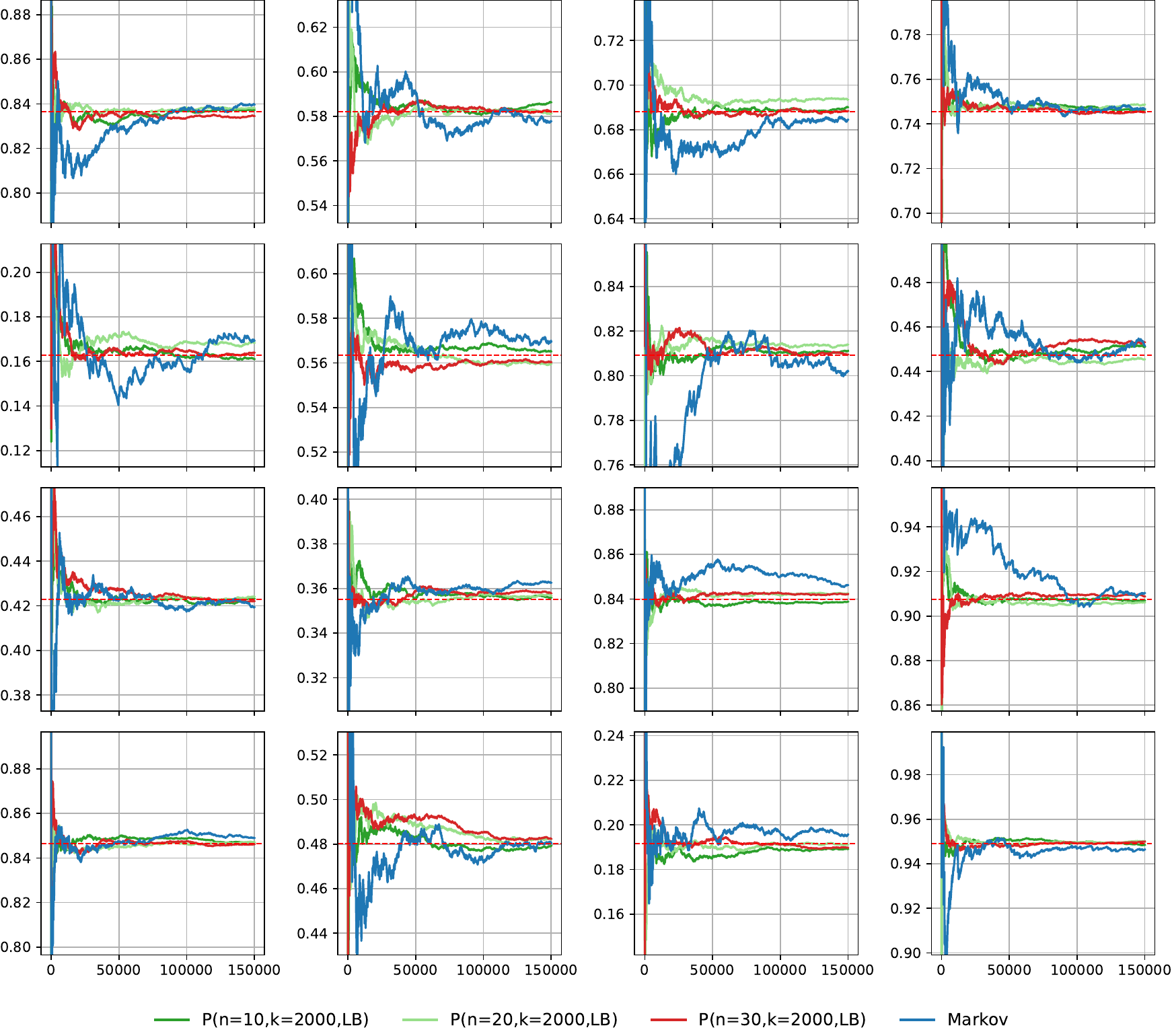}
}
    \caption{
        Number of steps to converge against true $p$-value (dotted) for $\mathrm{QI}_{5\times 5}$
    using basic moves only.}\label{fig:qi5x5_no_markov_basis}
\end{center}
\end{figure}

\section{Chances, risks and limitations}

The SAT based perspective offers several chances for conditional inference on contingency
tables. Most importantly, it enables methods that do not rely on Markov bases, which removes a major
computational barrier in models with many constraints and structural zeros. Modern SAT samplers
can accelerate practical convergence by moving quickly across
distant regions of the fiber, which reduces dependence between successive draws compared with purely
local moves. When a lattice basis connects the fiber within slices, it can be used to
generate simple proposals with low overhead, providing a practical alternative in settings where a
full Markov basis is unknown or too costly. These advantages are most pronounced in involved
examples with small sample sizes or a large number of structural zeros.
In these cases, we think our method closes a particular performance gap, while other methods like Heat-bath
based Markov chains~\cite{heat_bath_fiber_walks} that dynamically adjust the lengths of the moves
pay out their full potential in the limit of large sample sizes.

These gains come with risks that require careful mitigation. SAT samplers may introduce structural
sampling errors which can distort the empirical distribution on the fiber and bias conditional
$p$-values. Our hybrid-methods take care of these biases, but they also come with hyperparameters
one may need to tune, like the number of walks or steps per walk.
Finally, there are intrinsic limitations. The computation time per sample can be higher than for a
single local move because sampling often involves repeated solver calls, for example with XOR
constraints. However, the longer time can be a good investment, as subsequent
samples from the SAT-sampler may come from fiber regions far apart where subsequent samples of
Markov moves coincide in most of their entries.

For large design matrices and rich constraint sets, the Boolean
encoding can grow rapidly, which increases memory usage and may render the SAT instance too large
for practical solving. In such cases simpler lattice or Markov walks without SAT support or
asymptotic approximations may be preferable for very large fibers, while exact extraction can
dominate for very small fibers. Our implementation therefore includes fallbacks and tuning options,
such as slice granularity and restart policies, to navigate these regimes, but it does not remove
the fundamental scaling limits that arise from very high dimensional constraint systems.

\subsection*{Acknowledgements}
The authors gratefully acknowledge the support received from the
Hightech Agenda Bayern. We thank anonymous reviewers for their helpful comments and suggestions.

\subsection*{Data availablility}
The implemented algorithms are freely available under
\url{https://github.com/scharpfenecker/sat_fiber_walks}.

\subsection*{Conflict of interests}
The authors provide no conflict of interest associated with the content of this article.

\bibliographystyle{plainnat}
\bibliography{satwalks.bib}

\appendix

\section{Convergence of $\hybridAlg$}\label{app:bias}

In this section, we consider the bias and stationary distribution of the hybrid algorithm
$\hybridAlg$ when the SAT-sampler is only approximately uniform.
Throughout, $\cM$ is a Markov basis for $A$, so that
the local kernel $\cW_{\mathrm{MB}}$ is ergodic and reversible with respect to the target $\rho$
on $\fiber{A}{b}$, and $p_{\mathrm{SAT}}(u)$ denotes the probability that the SAT-sampler returns the solution
$u\in\fiber{A}{b}$, as in Section~\ref{s:sat_sampling}.

As a SAT-proposal $v\in\fiber{A}{b}$ which is drawn from $p_{\mathrm{SAT}}$ is independent of the current state $u$, the exact
Metropolis-Hastings acceptance probability is
$$\min\!\left(1,\,\frac{\rho(v)\,p_{\mathrm{SAT}}(u)}{\rho(u)\,p_{\mathrm{SAT}}(v)}\right),$$
but the marginals 
$p_{\mathrm{SAT}}(u)$ and $p_{\mathrm{SAT}}(v)$
of a SAT-sampler are intractable. As in our experiments we drop the
unknown ratio 
$\frac{p_{\mathrm{SAT}}(u)}{p_{\mathrm{SAT}}(v)}$ and accept with the simplified probability
$\min(1,\frac{\rho(v)}{\rho(u)})$. Writing the resulting effective SAT-kernel
for $u\neq v$ as
\[
    \widetilde{\cW}_{\mathrm{SAT}}(u,v)=p_{\mathrm{SAT}}\cdot (v)\,\min\!\left(1,\frac{\rho(v)}{\rho(u)}\right),
\]
with the usual rejection mass on the diagonal. The next lemma shows its stationary distribution:

\begin{lemma}\label{lem:sat_target}
    The stationary distribution of the Markov chain induced by
$\widetilde{\cW}_{\mathrm{SAT}}$ is
$$\tilde p_{\mathrm{SAT}}(u)\sim \rho(u)\,p_{\mathrm{SAT}}(u),$$

\end{lemma}

\begin{proof}
    Let $Z=\sum_{u\in\fiber{A}{b}}\rho(u)\cdot p_{\mathrm{SAT}}(u)$ be the normalising constant.
For any $u\neq v$ we have the identity
$$\frac{\min(1,\rho(v)/\rho(u))}{\min(1,\rho(u)/\rho(v))}=\frac{\rho(v)}{\rho(u)}.$$
Assume without loss of generality $\rho(v)\geq\rho(u)$, so that
$\min(1,\rho(v)/\rho(u))=1$ and $\min(1,\rho(u)/\rho(v))=\rho(u)/\rho(v)$. Then
\[
  \tilde p_{\mathrm{SAT}}(u)\cdot \widetilde{\cW}_{\mathrm{SAT}}(u,v)
  =\frac{\rho(u)\cdot p_{\mathrm{SAT}}(u)}{Z}\cdot p_{\mathrm{SAT}}(v)\cdot 1
  =\frac{\rho(u)\cdot p_{\mathrm{SAT}}(u)\cdot p_{\mathrm{SAT}}(v)}{Z},
\]
while
\[
  \tilde p_{\mathrm{SAT}}(v)\cdot \widetilde{\cW}_{\mathrm{SAT}}(v,u)
  =\frac{\rho(v)p_{\mathrm{SAT}}(v)}{Z}\cdot p_{\mathrm{SAT}}(u)\cdot\frac{\rho(u)}{\rho(v)}
  =\frac{\rho(u)\cdot p_{\mathrm{SAT}}(u)\cdot p_{\mathrm{SAT}}(v)}{Z}.
\]
As these two quantities agree, the detailed balance equation holds
and $\tilde p_{\mathrm{SAT}}$ is the stationary distribution of $\widetilde{\cW}_{\mathrm{SAT}}$.
\end{proof}

The bias a $\epsilon$-uniform SAT-sampler introduces is therefore \emph{systematic}: it reflects the
sampler's structural non-uniformity and cannot be removed by running the chain longer. Its
magnitude is controlled directly by the uniformity parameter $\epsilon$.

\begin{lemma}\label{lem:sat_eps}
Suppose the SAT-sampler is $\epsilon$-uniform, i.e.\
$\tfrac{1-\epsilon}{\lvert\fiber{A}{b}\rvert}\leq p_{\mathrm{SAT}}(u)\leq \tfrac{1+\epsilon}{\lvert\fiber{A}{b}\rvert}$
for all $u\in\fiber{A}{b}$ and some $\epsilon\in[0,1)$, then
$$\bigl\lVert \tilde p_{\mathrm{SAT}}-\rho\bigr\rVert_{\mathrm{TV}}
  \;\leq\;\frac{\epsilon}{1-\epsilon}.$$
\end{lemma}
\begin{proof}
By Lemma~\ref{lem:sat_target}, we have
$$\frac{\tilde p_{\mathrm{SAT}}(u)}{\rho(u)}=\frac{p_{\mathrm{SAT}(u)}}{Z},$$
    with normalizing constant $Z=\sum_w\rho(w)p_{\mathrm{SAT}}(w)$.
    Clearly, as $p_{\mathrm{SAT}}$ is $\epsilon$-uniform and $\sum_{w}\rho(u)=w$, we have

$$\tfrac{1-\epsilon}{\lvert\fiber{A}{b}\rvert}\leq Z \leq
\tfrac{1+\epsilon}{\lvert\fiber{A}{b}\rvert}$$
as well. Dividing the bounds on $p_{\mathrm{SAT}}(u)$ by those on $Z$ gives for each $u\in\fiber{A}{b}$:
\[
  \frac{1-\epsilon}{1+\epsilon}\leq\frac{\tilde p_{\mathrm{SAT}}(u)}{\rho(u)}\leq\frac{1+\epsilon}{1-\epsilon}
\]
This gives
\[
  \bigl\lVert \tilde\rho_{\mathrm{SAT}}-\rho\bigr\rVert_{\mathrm{TV}}
  =\tfrac12\sum_{u\in\fiber{A}{b}} \rho(u)\,\Bigl\lvert\tfrac{p_{\mathrm{SAT}}(u)}{Z}-1\Bigr\rvert
  \leq \tfrac12\max_{u\in\fiber{A}{b}}\Bigl\lvert\tfrac{p_{\mathrm{SAT}}(u)}{Z}-1\Bigr\rvert
  \leq \tfrac12\Bigl(\tfrac{1+\epsilon}{1-\epsilon}-1\Bigr)
  =\frac{\epsilon}{1-\epsilon}.
\]
\end{proof}

Of course, the hybrid scheme $\hybridAlg$ does not iterate $\widetilde{\cW}_{\mathrm{SAT}}$ alone: each
SAT-step is interleaved with $n-1$ exact local steps, so one block of the chain is governed by the
kernel
\[
  P_n=\cW_{\mathrm{MB}}^{\,n-1}\,\widetilde{\cW}_{\mathrm{SAT}}.
\]
Because $\cW_{\mathrm{MB}}$ is $\rho$-reversible, the local steps preserve $\rho$ exactly and only
\emph{pull} the distribution back towards $\rho$; the SAT-step is the sole source of bias. The next
result quantifies how diluting the SAT-steps (large $n$) suppresses the residual bias of the
composite chain. 
As stated in Section~\ref{s:hybrid}, we measure the distance of the stationary distributions using
the
Dobrushin ergodic coefficient of the $(n-1)$-step kernel,
\[
  c_{n-1}=c(\cW_{\mathrm{MB}}^{\,n-1})=\sup_{u,v}\bigl\lVert
  \cW_{\mathrm{MB}}^{n-1}(u,\cdot)-\cW_{\mathrm{MB}}^{n-1}(v,\cdot)\bigr\rVert_{\mathrm{TV}}\in[0,1].
\]
With this, we now can prove the main statement from Section~\ref{s:hybrid}:
\begin{proof}[Proof of Theorem~\ref{thm:hybrid_bias}:]

We write
  $\cW=\cW_{\mathrm{MB}}^{\,n-1}\,\widetilde{\cW}_{\mathrm{SAT}}$ and define
$d=\lVert\rho_n-\rho\rVert_{\mathrm{TV}}$. Moreover, we define the distribution
$\mu=\rho_n^T\cdot\cW_{\mathrm{MB}}^{n-1}$ on the fiber.
As $\rho^T\cdot\cW_{\mathrm{MB}}^{n-1}=\rho$, we obtain
\[
  \lVert\mu-\rho\rVert_{\mathrm{TV}}=
  \|\rho_n^T\cdot\cW_{\mathrm{MB}}^{n-1}-\rho^T\cdot\cW_{\mathrm{MB}}^{n-1}\|_{\mathrm{TV}}=
  \bigl\lVert(\rho_n-\rho)^T\cW_{\mathrm{MB}}^{n-1}\bigr\rVert_{\mathrm{TV}}
  \leq d\cdot c_{n-1}.
\]

Clearly $\rho_n=\mu^T\cdot\widetilde{\cW}_{\mathrm{SAT}}$ as $\rho_n$ is by definition the
stationary distribution of $\cW$. Thus, we can write
\[
  d=\bigl\lVert\mu^T\widetilde{\cW}_{\mathrm{SAT}}-\rho\bigr\rVert_{\mathrm{TV}}
  \leq \bigl\lVert\mu^T\widetilde{\cW}_{\mathrm{SAT}}-\rho\widetilde{\cW}_{\mathrm{SAT}}\bigr\rVert_{\mathrm{TV}}
  +\bigl\lVert\rho^T\widetilde{\cW}_{\mathrm{SAT}}-\rho\bigr\rVert_{\mathrm{TV}}.
\]
We bound both summands separately from above. First, we use
  $$\bigl\lVert\mu^T\widetilde{\cW}_{\mathrm{SAT}}-\rho\widetilde{\cW}_{\mathrm{SAT}}\bigr\rVert_{\mathrm{TV}}
=\bigl\lVert(\mu-\rho)^T\cdot \widetilde{\cW}_{\mathrm{SAT}}\bigr\rVert_{\mathrm{TV}}
  \le
  \lVert\mu-\rho\rVert_{\mathrm{TV}}$$
  because a stochastic kernel is non-expansive in total variation distance. 
  For the second summand, we use the triangle inequality and non-expansiveness again to obtain
\[
  \bigl\lVert\rho\widetilde{\cW}_{\mathrm{SAT}}-\rho\bigr\rVert_{\mathrm{TV}}
  \leq \bigl\lVert\rho\widetilde{\cW}_{\mathrm{SAT}}-\tilde\rho_{\mathrm{SAT}}\widetilde{\cW}_{\mathrm{SAT}}\bigr\rVert_{\mathrm{TV}}
  +\bigl\lVert\tilde\rho_{\mathrm{SAT}}-\rho\bigr\rVert_{\mathrm{TV}}
  \leq 2\bigl\lVert\tilde\rho_{\mathrm{SAT}}-\rho\bigr\rVert_{\mathrm{TV}}.
\]
Putting everything together, we get $d\le d\cdot
c_{n-1}+2\lVert\tilde\rho_{\mathrm{SAT}}-\rho\rVert_{\mathrm{TV}}$ and because $c_{n-1}<1$ by
definition, we obtain:
$$d\leq \frac{2\lVert\tilde\rho_{\mathrm{SAT}}-\rho\rVert_{\mathrm{TV}}}{1-c_{n-1}}.$$
Using the bound from Lemma~\ref{lem:sat_eps} proves the claim.
\end{proof}

\section{SAT-encoding example}\label{app:sat_encoding}

For an example for encoding the fiber sampling as a SAT sampling problem, consider the fiber defined
by the following matrix $A$ and right-hand side $b$:
\[A = \begin{bmatrix}
1 & 2 \\
1 & 1
\end{bmatrix}, b = \begin{bmatrix}
3 \\
2
\end{bmatrix}.
\]
We can rewrite the fiber equation $Ax = b$ as a system of linear equations (one per line) and combine the equations with a boolean conjunction:
\[\begin{bmatrix}
1 & 2 \\
1 & 1
\end{bmatrix} \cdot \begin{bmatrix}
u_1 \\
u_2
\end{bmatrix} = \begin{bmatrix}
3 \\
2
\end{bmatrix} \Leftrightarrow 1 \cdot u_1 + 2 \cdot u_2 = 3 \land 1 \cdot u_1 + 1 \cdot u_2 = 2.
\]
Then the largest entry on the right side has $2$ bits, so $l = 2$, and all solutions to $Ax = b$ can be embedded in $[2^2 - 1]^2 = \{0, \ldots, 3\}^2$ and equivalently in $\{0, 1\}^4$ using binary encodings. 
As a boolean circuit, the system can be expressed as
\[
C = \texttt{equiv}(\texttt{add}(\texttt{mul}(1, u_1), \texttt{mul}(2, u_2)), 3) \land \texttt{equiv}(\texttt{add}(\texttt{mul}(1, u_1), \texttt{mul}(1, u_2)), 2).
\]

The constants get replaced by their binary encoding, $u_1$ and $u_2$ each are encoded by $l$ free
variables $u_{11}, u_{12}, u_{21}, u_{22}$ and for $\texttt{equiv}$, $\texttt{add}$ and
$\texttt{mul}$ we use appropriate boolean circuits performing these operations.

Writing each entry from most to least significant bit as $u_i = 2\,u_{i2} + u_{i1}$, the constants on the right-hand side
become $3 = (1,1)$ and $2 = (1,0)$. The multiplications by the matrix entries are particularly
simple: $\texttt{mul}(1,u_i) = u_i$ is the identity, while $\texttt{mul}(2,u_2)$ is a left shift,
yielding the three-bit number $(u_{22}, u_{21}, 0)$ that represents $2u_2$. In the more general case, multiplication schemes like full adders, naive long multiplication or e.g. karatsuba multiplication can be used, especially when both inputs are non-constant.

Consider the first equation $u_1 + 2u_2 = 3$. Adding $u_1 = (0, 0, u_{12}, u_{11})$ and
$2u_2 = (0, u_{22}, u_{21}, 0)$ with a ripple-carry adder (a chain of full adders, lowest bit first)
produces the sum bits
\[
\begin{aligned}
s_0 &= u_{11}, &
s_1 &= u_{12} \oplus u_{21}, &
s_2 &= u_{22} \oplus (u_{12} \land u_{21}), &
s_3 &= u_{12} \land u_{21} \land u_{22},
\end{aligned}
\]
where each $\oplus$, $\land$ and intermediate carry is one gate of the circuit $C$. The wrapping
$\texttt{equiv}(\,\cdot\,, 3)$ compares this sum bit-by-bit against $3 = (0, 0, 1, 1)$, i.e. it enforces
$s_0 = 1$, $s_1 = 1$, $s_2 = 0$ and $s_3 = 0$. The second equation $u_1 + u_2 = 2$ is encoded
analogously, summing $u_1 = (0, u_{12}, u_{11})$ and $u_2 = (0, u_{22}, u_{21})$ to
\[
\begin{aligned}
t_0 &= u_{11} \oplus u_{21}, &
t_1 &= u_{12} \oplus u_{22} \oplus (u_{11} \land u_{21}), &
t_2 &= \mathrm{maj}\bigl(u_{12}, u_{22}, u_{11} \land u_{21}\bigr),
\end{aligned}
\]
and comparing against $2 = (0,1,0)$, i.e. $t_0 = 0$, $t_1 = 1$, $t_2 = 0$.

Applying the Tseitin transformation, every gate is replaced by its defining clauses. For instance,
the carry $c = u_{12} \land u_{21}$ appearing in both $s_2$ and $s_3$ is captured by a fresh variable
$c$ together with
\[
(\neg c \lor u_{12}) \land (\neg c \lor u_{21}) \land (c \lor \neg u_{12} \lor \neg u_{21}),
\]
and the $\oplus$-gates are encoded by the analogous four-clause pattern. Conjoining the clauses of
all gates with the unit clauses that fix the output bits to the binary digits of $b$ yields the CNF
formula $F$ over the four free variables $u_{11}, u_{12}, u_{21}, u_{22}$ and the gate-helper
variables. By construction $F$ is satisfiable if and only if $\fiber{A}{b} \cap ([2^l-1])^d$ is
nonempty, and its satisfying assignments are in bijection with the fiber elements.

Here the system has the single solution $u_{11} = 1$, $u_{12} = 0$, $u_{21} = 1$, $u_{22} = 0$, which
decodes to $u = (1,1)$, so $F$ has a unique satisfying assignment and the fiber
$\fiber{A}{b} = \{(1,1)\}$ is a singleton. For a fiber with more than one element the same
construction produces one satisfying assignment per table, and a SAT sampler draws among them.

\section{Guidance on local move set}

To test the effect of the local move set $\cM$ on the convergence, we run experiments in
$\mathrm{I}_{5\times 5}$ for two initial observations:

$$u^{\mathrm{obs}}_1 = \begin{bmatrix}
0& 0& 1& 0& 0\\
0& 0& 0& 0& 1\\
1& 3& 0& 0& 1\\
0& 1& 0& 0& 0\\
0& 0& 0& 2& 0
\end{bmatrix}
\quad\text{
and}\quad u^{\mathrm{obs}}_2 = \begin{bmatrix}
0& 0& 0& 1& 0 \\
1& 0& 0& 0& 0 \\
0& 4& 0& 1& 0 \\
0& 0& 0& 0& 1 \\
0& 0& 1& 0& 1
\end{bmatrix}
$$
Both elements are in the same fiber, which can be enumerated completely using our SAT-encoding. We
selected these two observations as they have $p$-values slightly above and below the significance
level of $0.05$ respectively: $0.0683$ for $u^{\mathrm{obs}}_1$ and $0.0349$ for
$u^{\mathrm{obs}}_2$. In our experiments, we varied the number of local moves $\cM$, computed the
number of connected components in the fiber, and visualized the convergence to the true $p$-values.

\begin{figure}[!ht]
\begin{center}
    \subfloat[Convergences for independence tests for $u^{\mathrm{obs}}_1$ with $p$-value above
    significance level]{
    \includegraphics[width=\textwidth]{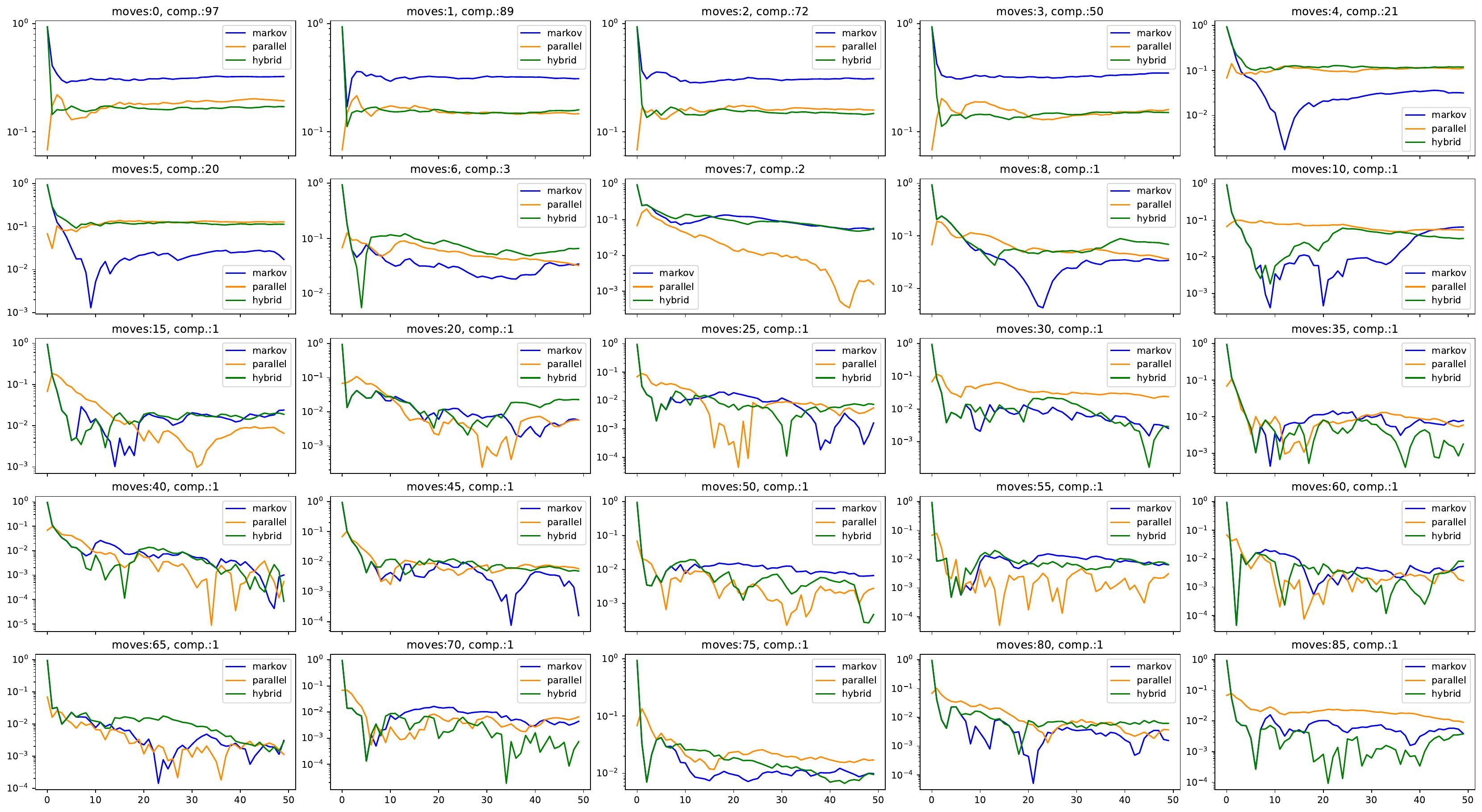}
}\\
\subfloat[Convergences for independence tests for $u^{\mathrm{obs}}_1$ with $p$-value above
    significance level]{
        \includegraphics[width=\textwidth]{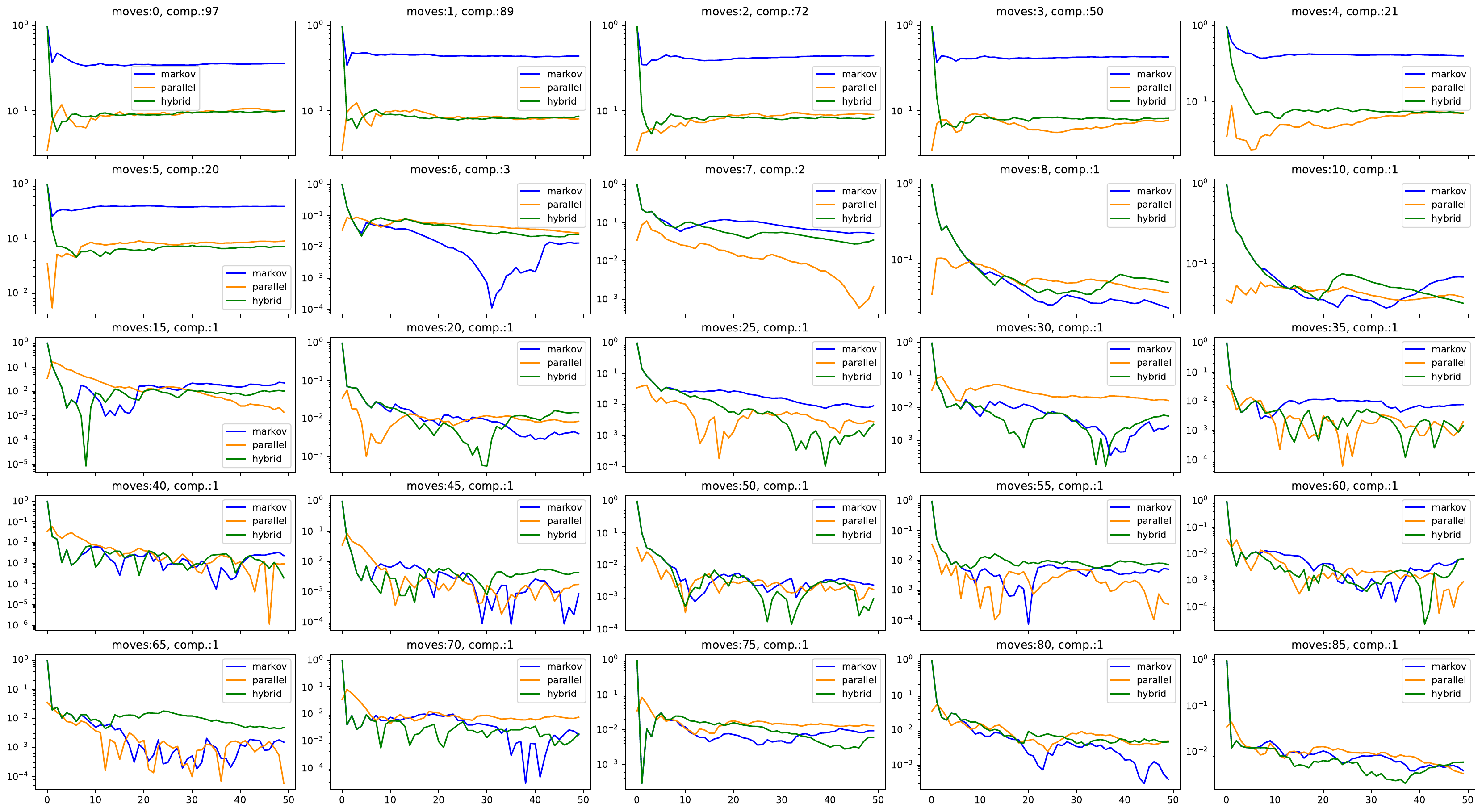}
    }
    \caption{
        Convergence rates for different set of moves averaged over $7$ runs.}\label{fig:non_markov_convergence}
\end{center}
\end{figure}

Figure~\ref{fig:non_markov_convergence} shows the estimation error for the true $p$-value as a
function of the number of sampling steps (in thousands), for a growing number of local moves $\cM$ (see moves in title). The moves are initially only the Lattice basis, e.g. moves:0 adds no additional moves until moves:85 adds all remaining moves for the Markov basis. When $\cM$ does
not connect the fiber (component count above one), the pure local Markov walk stagnates at a biased
estimate, since it never leaves its initial component, while both the parallel and the hybrid
SAT-schemes manage more often to converge to the true $p$-value. As $|\cM|$ increases the number of components drops rapidly to one --- at which point
$\cM$ is a Markov basis --- and all three methods converge at comparable rates, with the additional
local moves accelerating the mixing of the hybrid chain. The behaviour is similar for the
observation above ($p=0.0683$) and below ($p=0.0349$) the $0.05$ significance level, so the
SAT-augmented samplers recover the correct test decision even in regimes where the local moves alone
do not.

\section{Metropolis-Hastings algorithm}\label{app:mh}

\begin{algorithm}[H]
    \caption{Metropolis-Hastings for $p$-value approximation}
    \SetKwInOut{Input}{Input}
    \SetKwInOut{Output}{Output}
    \Input{initial observation $u^{\mathrm{obs}}$, number of steps $N$, random walk $\cW$, extremness measure $X$}
    \Output{Sequences $p\in[0,1]^N$}

    Set $u_0=u^{\mathrm{obs}}$
    \For{\(i \in [N]\)}{
    Sample $v\sim\cW(u_{i-1},\cdot)$\\
    With probability $\reject{\cW}{u_{i-1}}{v}$, set $u_i=v$, otherwise $u_i=u_{i-1}$\\
    Compute $p_i=\frac{1}{i}\sum_{j=1}^i\mathbf{1}_{X(u_j)\ge X(u^{\mathrm{obs}})}$
    }
    \Return{$p$}
    \label{alg:mh}
\end{algorithm}

\section{Further Results}\label{a:results}

\begin{figure}[!ht]
\begin{center}
    \includegraphics[height=0.18\textheight]{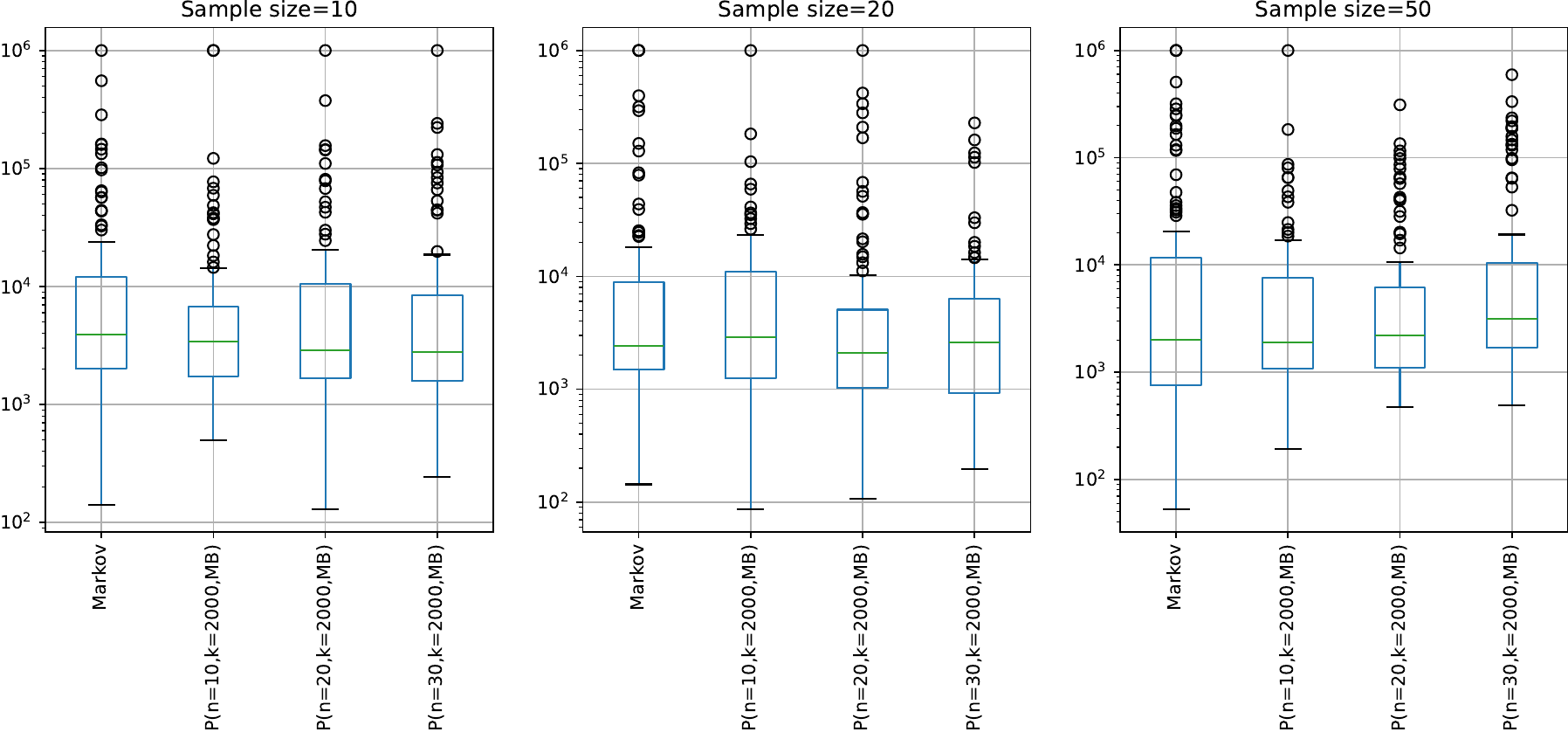}
    \caption{Results for $\mathrm{I}_{5\times 5}$.}
\end{center}
\end{figure}

\begin{figure}[!ht]
\begin{center}
    \includegraphics[height=0.18\textheight]{QI_5x5_terminations_0005.pdf}
    \caption{Results for $\mathrm{QI}_{5\times 5}$}
\end{center}
\end{figure}

\begin{figure}[!ht]
\begin{center}
    \includegraphics[height=0.18\textheight]{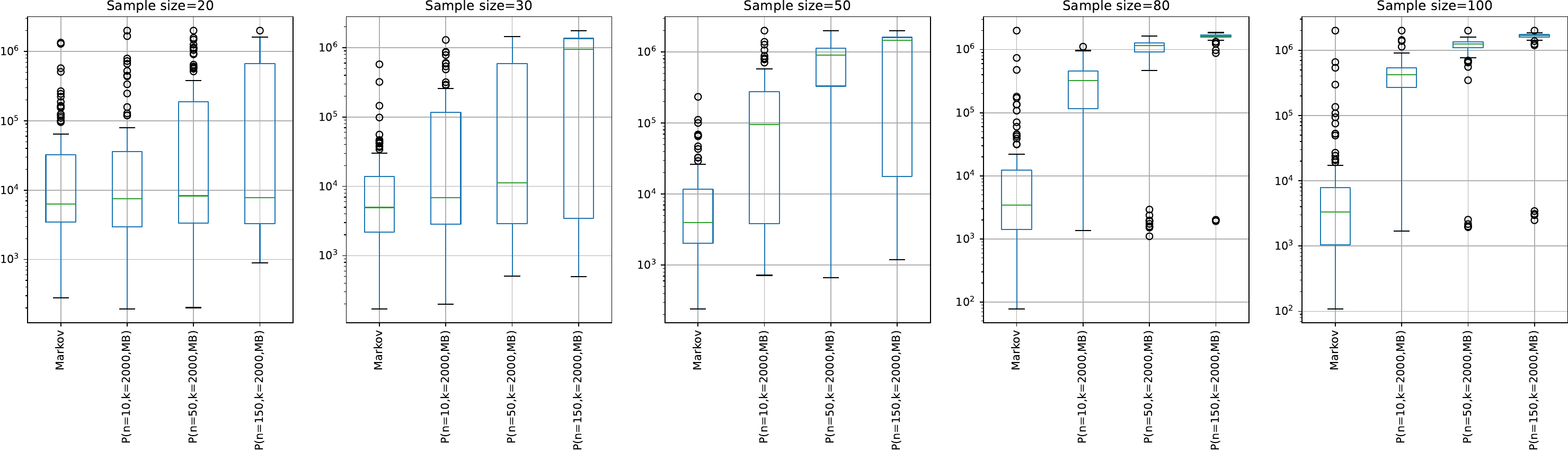}
    \caption{Results for $\mathrm{I}_{10\times 10}$.}
\end{center}
\end{figure}

\begin{figure}[!ht]
\begin{center}
    \includegraphics[height=0.18\textheight]{QI_10x10_terminations_0005.pdf}
    \caption{Results for $\mathrm{QI}_{10\times 10}$.}
\end{center}
\end{figure}

\begin{figure}[!ht]
\begin{center}
    \subfloat[$d=3$]{\includegraphics[height=0.18\textheight]{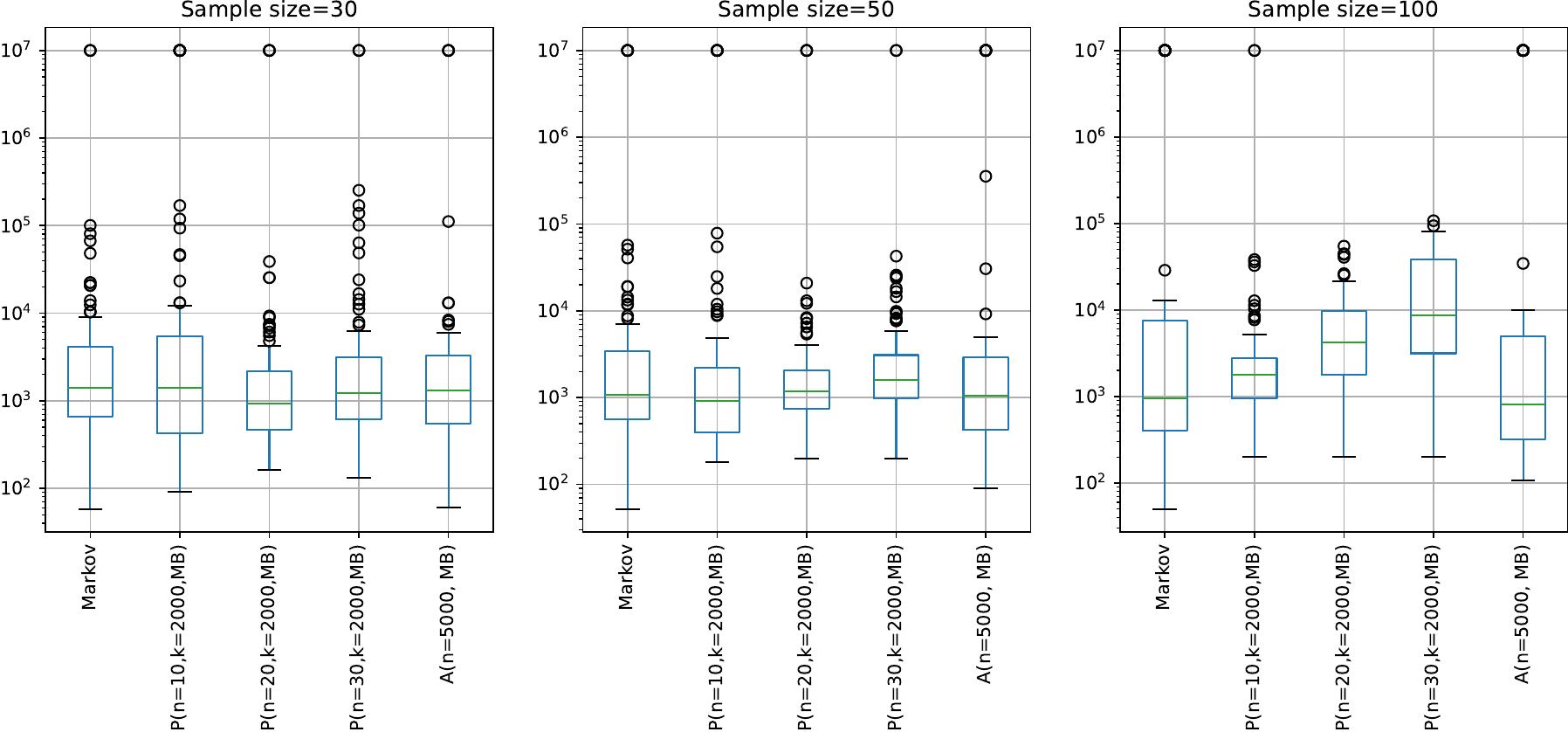}}
    \hfill
    \subfloat[$d=4$]{\includegraphics[height=0.18\textheight]{N3F_4_terminations_0005.pdf}}
    \caption{Results for $\mathrm{N3F}_{d}$.}
\end{center}
\end{figure}

\begin{figure}[!ht]
\begin{center}
    \subfloat[$k=4$]{\includegraphics[width=0.30\textwidth]{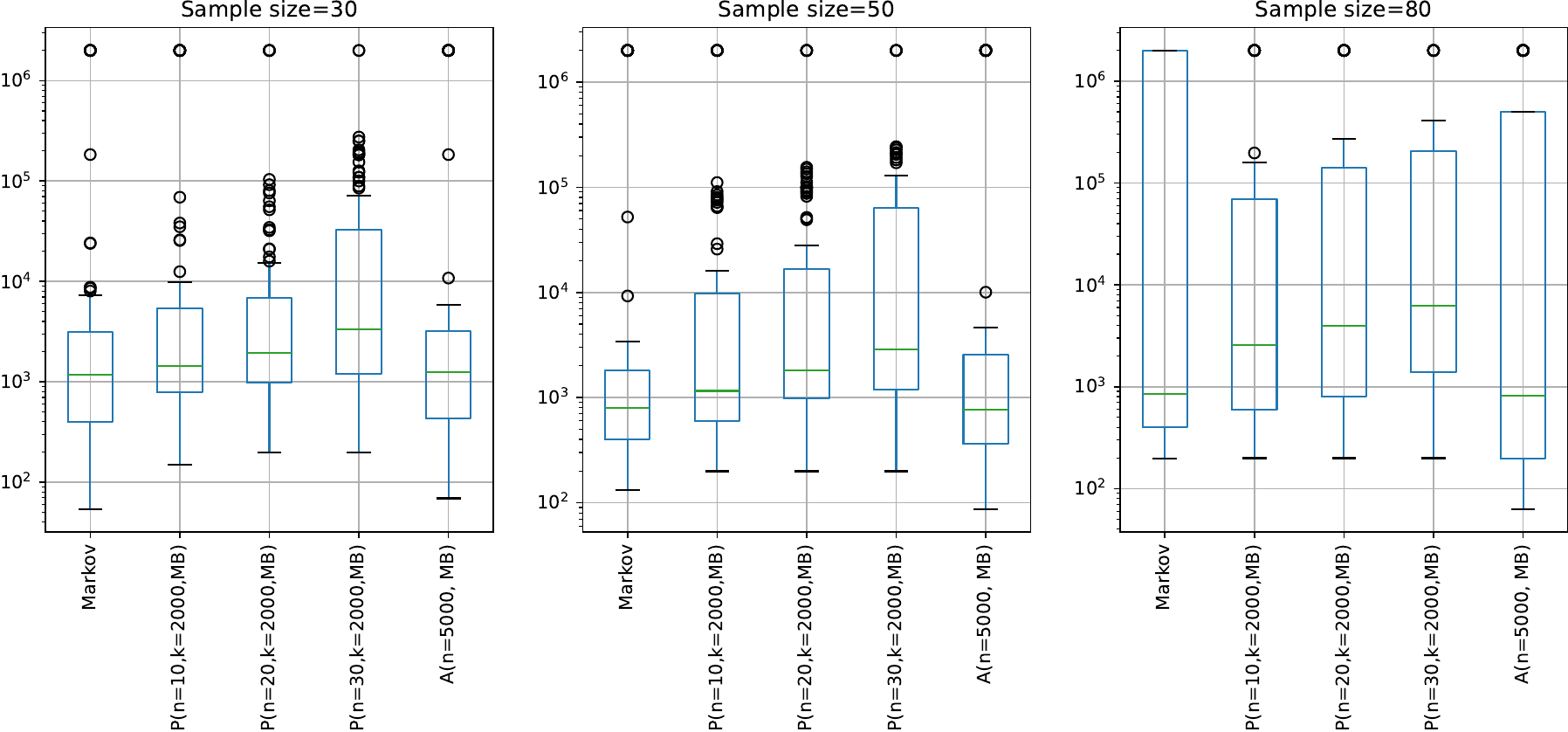}}
    \subfloat[$k=5$]{\includegraphics[width=0.30\textwidth]{I_5x2x2_terminations_0005.pdf}}
    \subfloat[$k=6$]{\includegraphics[width=0.30\textwidth]{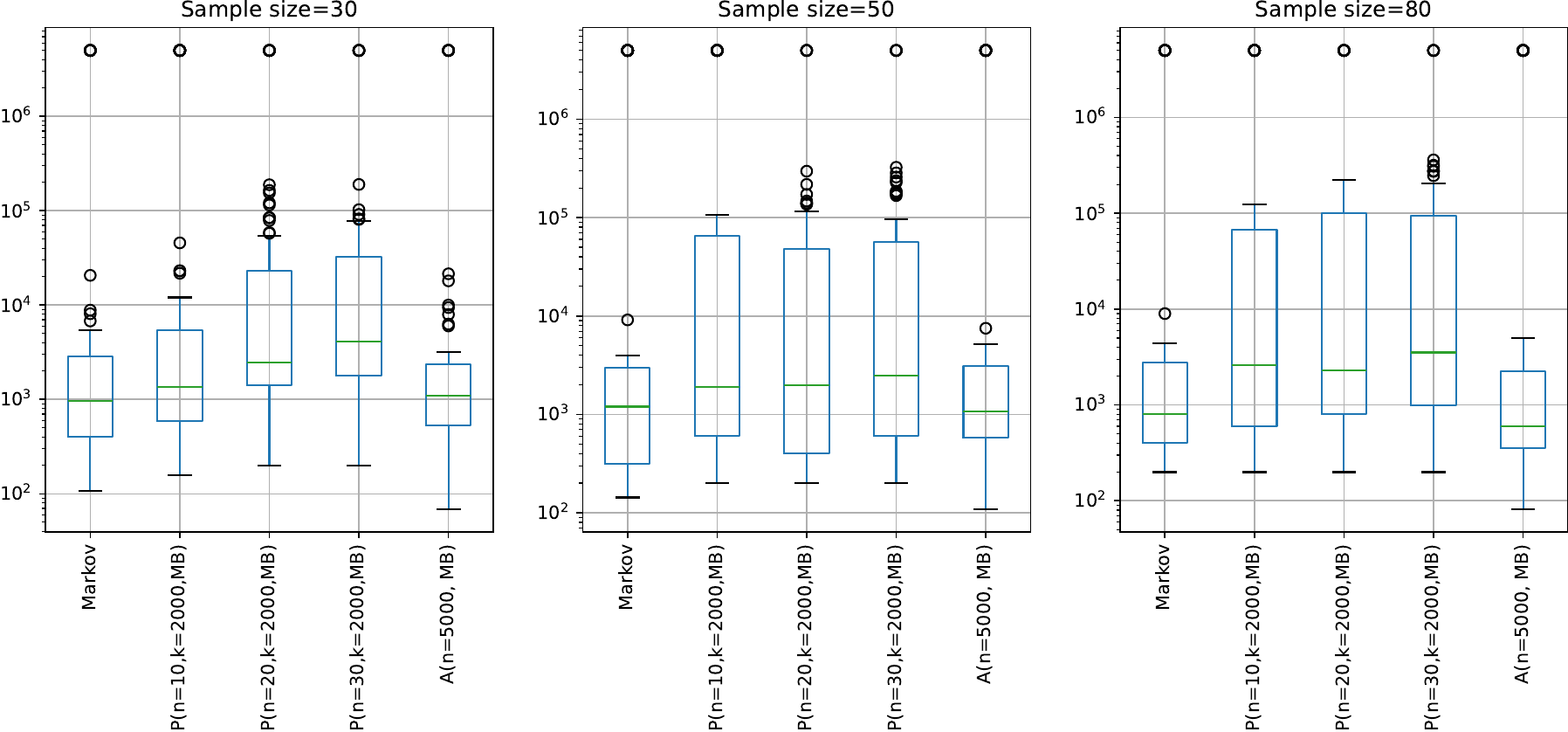}}
    \caption{Results for $\mathrm{I}_{k\times 2\times 2}$.}\label{fig:ik22}
\end{center}
\end{figure}

\end{document}